\documentclass{article}
\usepackage[utf8]{inputenc}
\usepackage[british]{babel}
\usepackage[T1]{fontenc}
\usepackage{graphicx,array}
\usepackage{caption}
\usepackage{subcaption}

\usepackage{authblk}
\usepackage{hyperref}
\usepackage{tikz}
\usetikzlibrary{calc}
\usepackage{authblk}
\usepackage{setspace}

\usepackage{cite}
\usepackage{float}

\usepackage{chemformula}
\usepackage{tikz}
\usetikzlibrary{positioning}
\usepackage{siunitx}
\usepackage{palatino}
\usetikzlibrary{external}
\usepackage{multirow}
\usepackage{dcolumn}
\usepackage{booktabs}
\usepackage{algorithmic}
\begin{document}
\title{A laser projection system for polymer transverse strain measurements in tensile testing}
\author[1,$^*$]{Dawid Kucharski}
\author[2]{Maciej Obst}
\author[3]{Jarosław Adamiec}
\affil[1]{Division of Metrology and Measurement Systems,
Institute of Mechanical Technology}
\affil[2]{Division of Strength of Materials and Structures,
Institute of Applied Mechanics}
\affil[3]{Institute of Machine Design}
\affil[ ]{Poznan University of Technology, Poland}
\affil[$^*$]{\href{mailto:dawid.kucharski@put.poznan.pl}{dawid.kucharski@put.poznan.pl}}
\maketitle
\doublespacing

\begin{abstract}
A self-built laser projection system for optical transverse strain measurements in tensile testing is presented. The setup based on laser diode modules, CCD (Charge-Coupled Device) array detectors, and standard optomechanical components is proposed to be a low-cost system for testing polymer samples. The optical setup is mobile and can easily be mounted on a tensile test machine. Circular cross-sectional changes in polymer diameter were detected with ten-micrometre resolution and stress-strain curves obtained by a tensile test machine.\\
The article describes signal processing using the open-source programming language \textsc{R}, and the transverse deformation of the polymer sample is evaluated in two perpendicular planes. The results obtained by the optical system may be used to accurately describe the models of energy-based mechanical properties of polymers for complex load conditions.


\end{abstract}
\section{Introduction}\label{sec:intro}
The stress-strain characteristics of materials supplemented with accurate transverse deformation measurements are essential for a more precise description of the material's mechanical properties. A material is considered non-classical if it violates, or at least does not fully satisfy, one or more of the assumptions on which classical continuum mechanics are based. An example is a deviation from Hooke's law for polymers (elastic nonlinearity, etc.) opposite to the metals as the classical materials. Especially modern non-classical materials require the development of appropriate mechanical properties and accurate approximation in a wide range of strains.\\  
The stress-strain curves of metallic materials do not always provide a clear mechanical description of the sample; the effect may even be more present with polymers. As S. Tu et al.~\cite{Tu2020} pointed out for metals, the flow stress in the post-neck region is essential to evaluate the hardening of the metal sheet. The effect is not observed in polymers, but accurate detection of the necking point is crucial for the mechanical evaluation of polymers during uniaxial tensile tests, for example, for the better plastic instability creep and relaxation evaluation process.\\
C. G'sell and J. Jones in~\cite{Gsell1979} introduced a relatively simple circumferential gauge specially designed to measure the neck diameters of polymer specimens. They determined the flow curves of polyvinyl chloride (PVC) and high-density polyethene (HDPE). More precise techniques are based on optics. F. Addiego et al. used a video system to accurately determine axial strain using fluorescent markers illuminated with an ultraviolet lamp~\cite{Addiego2006}. A typical optical technique is extensometry, also introduced by G'Sell et al.~\cite{GSell1992}. The method based on displacement measurements was applied for the description of the mechanical behaviour of PTFE (polytetrafluoroethylene) by L. Nunes et al.~\cite{Nunes2011}. The method has some severe limitations. Markers for video extensometry can modify the mechanical response of the tested materials. S.~Todros et al. introduced the problem in biomedical polymers~\cite{Todros2019}.\\
An exciting approach to optical methods for the mechanical evaluation of cylindrical PMMA shells (polymethylmethacrylate) was introduced by W. Liu and X. Zhang~\cite{Liu2020c}. They interposed the samples in monochromatic light from a green laser with power~$P=50$~mW.\\
S. Muhammad and P. Jar postulated that in HDPEs, necking in thinner specimens generated a more significant reduction in thickness direction but lower in width direction~\cite{Muhammad2011}. The data obtained by the extensometer suggested that resistance to neck propagation was highest for the thinnest specimens, as their neck growth speed was lowest and the flow stress highest. The authors used the data for the FE simulation to mimic the necking process. Our proposed bi-axial optical setup might verify these results with direct strain measurements.

In the Industry 4.0 age (The Fourth Industrial Revolution), where the development of smart factories and autonomous systems is now widely observed, optical instruments are game-changers in metrology and mechanical engineering~\cite{SI_units_2019}. In addition to surface measurements~\cite{Leach2011, Kucharski_D._A_2020}, optics also provides sub-micrometre resolution and high precision in the evaluation of mechanical properties. L. Zhang et al. proposed a simple laser setup to monitor the necking process during a material tensile test~\cite{Zhang1998}. The setup was not finally constructed but presented a promising application also for polymers.\\
The projection technique is another group of optical methods for specimen geometry analysis during tensile testing. R. Srinivasan et al. presented the projection method based on the modified shadow moire technique to measure large and small structures. They discussed a single projector, a single exposure method for neck development detection in tensile testing~\cite{Srinivasan1982}. H. Hoffmann and C. Vogl used an optical projection scanner based on structured light to measure the specimen's geometry in the necking area~\cite{Hoffmann2003}.\\
An exciting study was presented by J. Ye et al.~\cite{Ye2015} on the plastic instability of HDPE deformed under tension. Using full-field 3D-DIC (Digital Image Correlation) and extensometry, they pointed out that a video-extensometer based on marker tracking cannot efficiently measure the strain rate in the specimen centre at significant strain levels when the markers are very deformed. This limitation can be easily overcome with our proposed no-marker setup. The most reliable way to study the rheological properties of materials is by direct measurement of the neck profile.\\
An interesting optical setup for measuring edge position was described by J. Fisher and T. Radil~\cite{Fischer2003}. They compared a point light source shadow projection system with commonly used methods. They pointed out that when the collimated light beam illuminated the measured object, the method did not need complex signal processing, and the measured object's edge position was directly equal to the position of the edge in the illumination profile. The resolution of the presented position was approximately $1$ $\mu m$. No practical application was shown.\\
A reasonably innovative method to visualise the initiation and propagation of the neck of polymer films during tensile deformation was proposed by S.~Kato et al.~\cite{Kato2021}. They used chemical modifications of the polymer by tetraarylsuccinonitrile (TASN) moieties and observed yellow fluorescence with electron paramagnetic resonance (EPR) spectroscopy. They evaluated the microforces generated in the amorphous regions during the uniaxial stretching of the crystalline polymers.\\
W. Ren et al. introduced a method to measure the response to the strain of electroactive polymers by Doppler laser interferometry~\cite{Ren2004}. They investigated Maxwell stress actuators made of silicone and thermoplastic polyurethane.

Polymers are widely used in many applications where the correct evaluation of mechanical properties is crucial. An exciting area is polymer-based optical sensing~\cite{Leal-Junior:18}. A. Leal-Junior et al.~\cite{Leal-Junior2018} described a temperature sensor based on polymer optical fibre (POF) where the fibre stress changes with temperature variation. The authors characterised the variation of the POF mechanical property using DMA (dynamic mechanical analysis). They proposed an analytical model and simulation for the evaluation of sensor behaviour. Validation of the model in temperature tests with the fibre sensor was provided. In~\cite{Leal-Junior2018a} A. Leal-Junior et al. proposed optical fibre sensing based on the dynamic mechanical analysis (DMA) of the polymer to obtain the Young modulus concerning the variation of strain, temperature, humidity and frequency. Polymer optical fibre sensors can also be used for transverse force measurements. A. Leal-Junior et al.~\cite{Leal-Junior2018c} described an interesting experimental setup for the characterisation of the strain and transverse force of fused and non-fused POFs.\\ 
Another quite exciting polymer application was presented by A. Leal-Junior et al.~\cite{Leal-Junior2018b} about 3D printed elements to obtain low-cost and custom-made sensing elements.\\
Polymer-based additive manufacturing also requires precise measurement tools to evaluate the mechanical properties of components. The self-built optical transverse strain measurement system proposed in the paper for the tensile testing of polymers might be a recommended support tool.

The paper proposes a simple, low-cost optical setup for fast and precise radius monitoring of polymer sample rods during a tensile test. The setup based on the bi-axial laser line projection system is proposed for the first time in the literature. The proposed construction is based on parallel beams and linear CCD (Charge-Coupled Device) detectors, where the position of the edge in the illumination profile is equal to the position of the measured object's edge. The metrology evaluation of the setup followed by the calibration of the steel rods is introduced. A self-prepared algorithm based on the \textsc{R} programming language~\cite{R_project} was used for signal processing and data evaluation. This paper presents tests of the optical setup on a tensile machine. The system is proposed as an additional measurement tool for the experimental verification of the mechanical properties models~\cite{Kurpisz2020}.\\
In the paper, the following are described:
\begin{itemize}
	\item the design and construction of the new optical system for fast and precise transverse strain monitoring of polymers in tensile testing;
	\item the fast open-source R programme developed for signal processing. The code is freely available on GitHub;
	\item metrology calibration estimation of the system;
	\item detection of the polymer sample size changes during the tensile tests on the machine for various types of polymer rods;
	\item different mechanical behaviour of the materials depends on the type and sample preparation;
	\item complex symmetrical and non-symmetrical mechanical effects observed by the sample radius monitoring;
	\item the importance of the direct sample strain measurements under the tensile testing or compressing, which the proposed low-cost system can achieve;
	\item the importance of further works with the setup to revise non-classical material mechanical properties models.
 \end{itemize} 

\section{Experimental details}\label{sec:exp_details}
\subsection{Setup}\label{subsec:setup}
\begin{figure}
      \centering
          \includegraphics[width=0.7\textwidth]{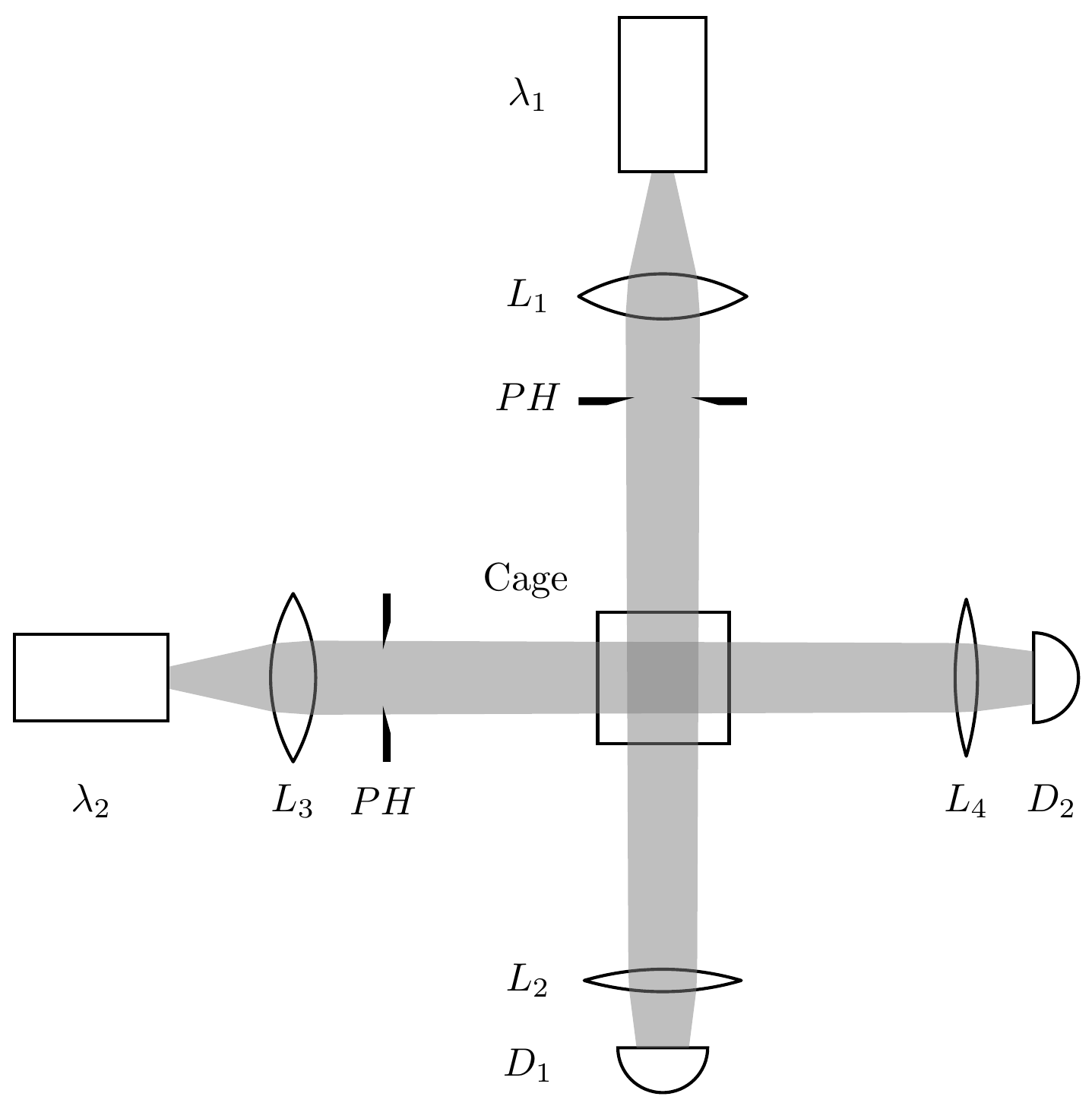}         
           \caption{Optical layout: $\lambda_{1,2}$ - semiconductor lasers ($\lambda=650$~nm, $P=5$~mW), $L_{1,3}$ - collimation lenses, $L_{2,4}$ - imaging lenses, $PH$ - pinhole, $D_{1,2}$ - CCD array detectors, $Cage$ - 30 mm cage cube}          
           \label{fig:setup}
      \end{figure}

The measurement system consists of two semiconductor lasers ($\lambda=650$~nm, $P=5$~mW), two CCD array detectors (TCD1304), two Nucleo boards (F401RE) and a self-built laser power supply. The optical setup is constructed with off-the-shelf optomechanical components. The optical layout of the system is shown in fig.~\ref{fig:setup}.
\begin{figure}
\centering
  \includegraphics[width=0.5\textwidth]{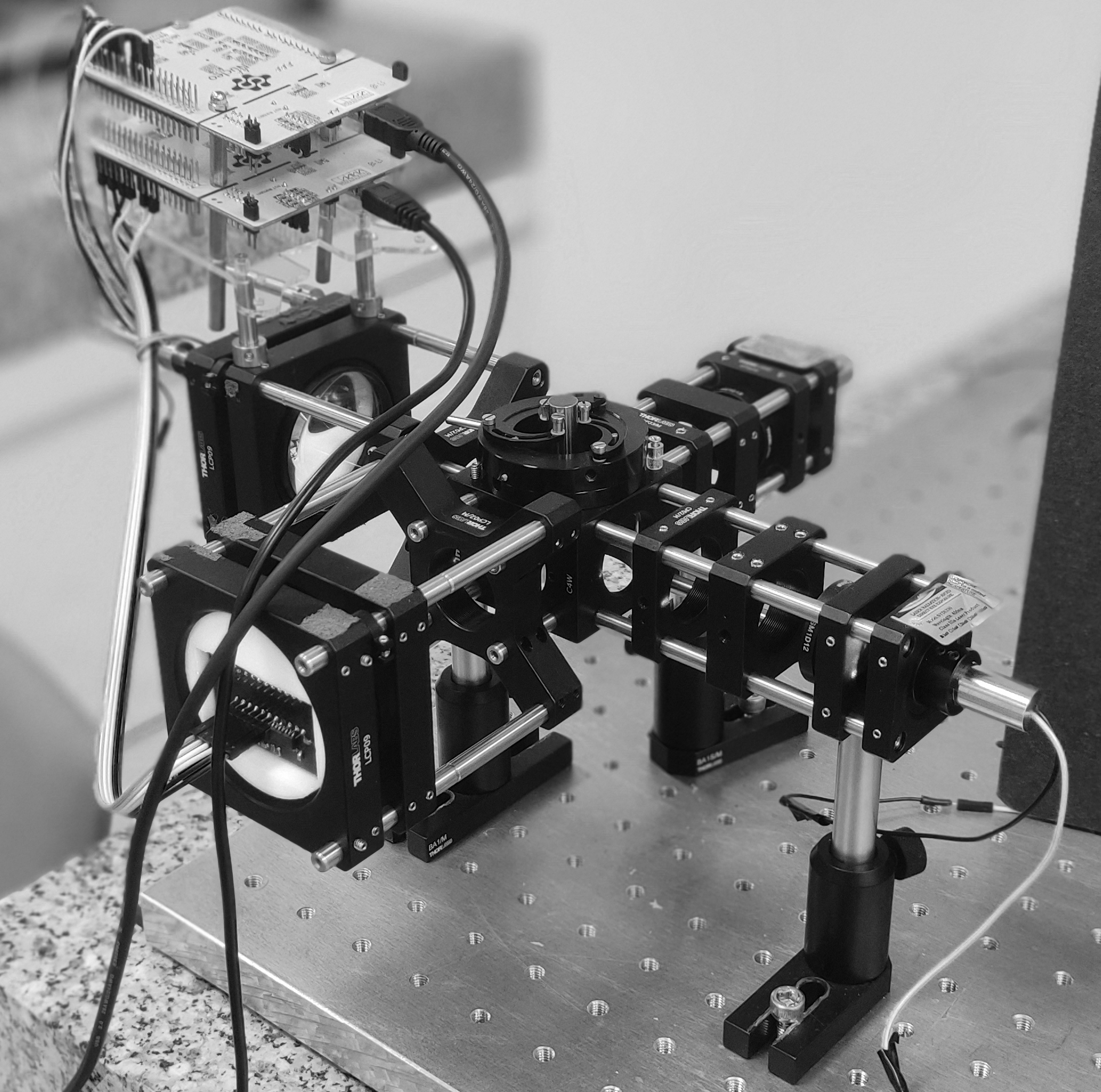}
  \caption{A perspective view of the setup}
  \label{fig:setup2}
\end{figure}
  
\begin{figure}
      \centering
      \begin{subfigure}[b]{0.48\textwidth}
          \centering
           \includegraphics[width=\textwidth]{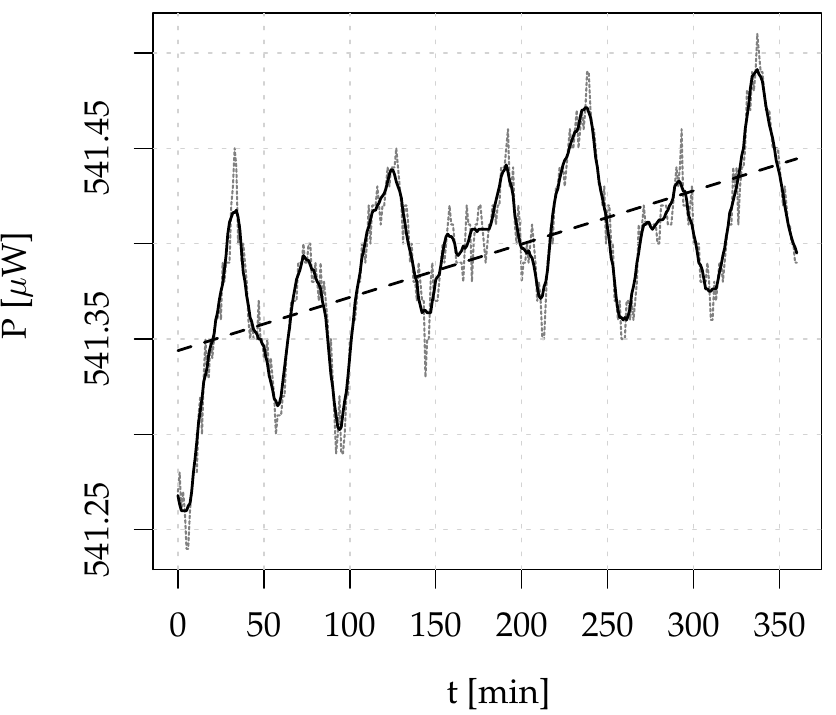}
           \caption[]%
           {{\small}}
           \label{fig:1diode_power}
      \end{subfigure}
      \hfill
      \begin{subfigure}[b]{0.48\textwidth}
          \centering
          \includegraphics[width=\textwidth]{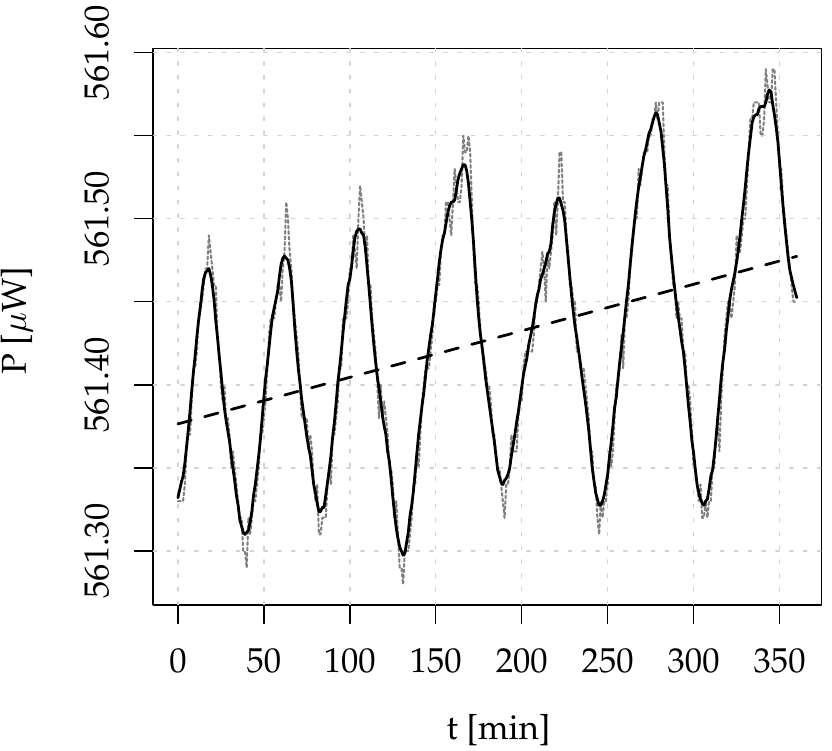}
         \caption[]%
          {{\small}}
          \label{fig:2diode_power}
      \end{subfigure}
      
      \begin{subfigure}[b]{0.48\textwidth}
          \centering
          \vskip\baselineskip
          \includegraphics[width=\textwidth]{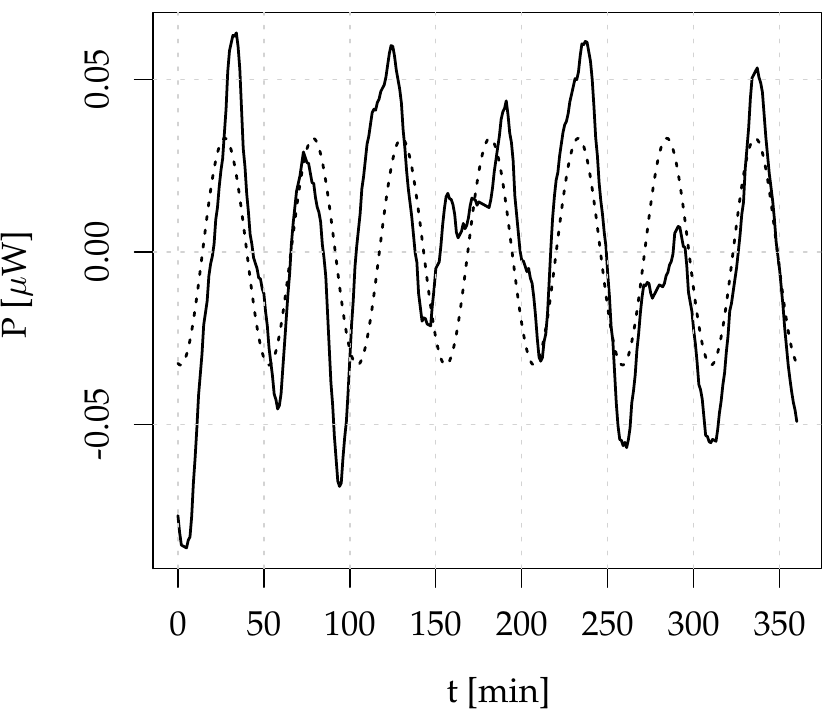}
          \caption[]%
          {{\small}}
          \label{fig:1diode_power_detrend}
      \end{subfigure}
      \hfill
      \begin{subfigure}[b]{0.48\textwidth}
          \centering
          \includegraphics[width=\textwidth]{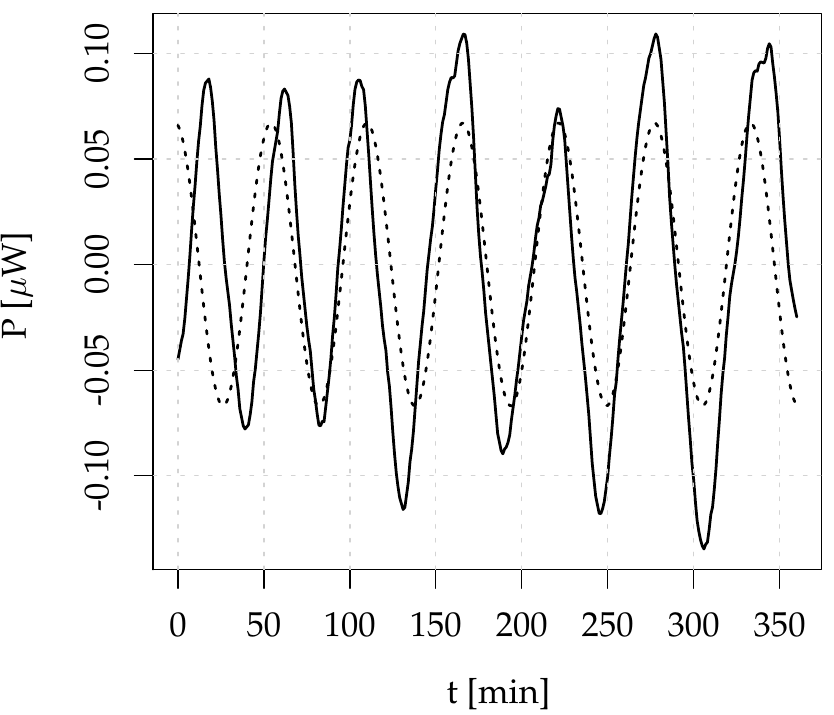}
          \caption[]%
          {{\small}}
    \label{fig:2diode_power_detrend}
      \end{subfigure}
      \caption[Lasers light power stability]
      {\small Lasers light power stability: (a) - oscillations of the $\# 1$ laser diode light power in time (grey dotted line: pure data, black line: average filtered data, dashed line: linear regression), (b) - oscillations of the $\# 2$ laser diode light power in time (grey dotted line: pure data, black line: average filtered data, dashed line: linear regression), (c) - linearly detrended oscillations of the $\# 1$ laser diode light power in time (black line: average filtered data, dotted line: cosine fitting curve), (d) - linearly detrended oscillations of the $\# 2$ laser diode light power in time (black line: average filtered data, dotted line: cosine fitting curve)}
      \label{fig:power_stability}
  \end{figure}

The light from the lasers is collimated into horizontal line-shaped beams by the lenses $L_{1,3}$. A sample is irradiated from two perpendicular directions for bi-axial radius detection. Two CCD array detectors ($D_1$ and $D_2$) register the shadow image during a test. In this way, real-time monitoring of the development of the sample neck can be performed. A self-built stabilised power supply ($I_{max}=150$~$A$; $V_{max}=30$~$V$; $V_{set}=4$~$V$) powered economic, non-stabilised lasers (no wavelength stabilisation). The lasers were tested for light-power constancy. Long-term linear drift ($\sim 6$ hours) and periodic oscillations were observed (see fig.~\ref{fig:1diode_power} and \ref{fig:2diode_power}). A moving average filter was used for the initial smoothing of the data. Linear and periodic detrending (removing a trend by model fitting) methods were implemented for optional final sample measurement results compensation if needed for time-consuming sample testing. To eliminate the linear trend, the \textsf{detrend} function of the \textsf{pracma} \textsc{R} package was applied~\cite{R_pracma}. For the elimination of periodic oscillations, a non-linear fitting function was used from the \textsf{stats} \textsc{R} package~\cite{R_project}. As shown in table~\ref{tab:power_stability}, the power instabilities of the lasers are approximately $0.30$ $\mu W$ and after trend reductions, less than $0.15$ $\mu W$ (estimated as the maximum amplitude.\\
The power instabilities are negligible for the application presented in the paper, for short-time measurements (approximately $10$ min. per sample), but it may be important for more advanced measurements for a time-consuming sub-micrometre application. The results presented in the paper with short-time measurements also avoid the second necking appearance problem in polymers tensile testing~\cite{Pink1989}. 
However, the laser parameters should be tested experimentally prior to use. It is crucial, especially for low-cost lasers ($\leq 1\pounds$ per diode), that might also be applied for an advanced purpose.
\begin{table}
\centering
\caption{Laser power instability estimations. $A$ -- Amplitude, $\omega$ -- frequency, $t$ -- time, $\phi$ -- phase, $C$ -- constant. The data correspond to fig.~\ref{fig:power_stability}}
\label{tab:power_stability}
\begin{tabular}{cccc}
\hline\hline
Detrending  & Function               & $\Delta P_{D_1}$ $[\mu W ]$ & $\Delta P_{D_2}$ $[\mu W ]$ \\ \hline
non      & -                      & $0.27$                & $0.31$                   \\ 
linear   & $y=a\cdot t+b$         & $0.15$                & $0.24$                   \\ 
periodic & $y=A\cdot\cos(\omega\cdot t+\phi)+C$ & $0.06$  & $0.13$                   \\ \hline
\end{tabular}
\end{table}
  
  The TCD1304 is a high-sensitivity 3648-pixel linear CCD. Two CCD detectors were applied in the setup. Two Nucleo boards (F401RE) were used for data registering, controlled by a command-line interface (CLI). The CCDs are analogue devices that collect photons and convert them into charge. The collected charge is then shifted to the output, where an analogue-to-digital converter (ADC) digitises it. The ADC signal is cleaner than the raw output when a capture is done with the ADC of the microcontroller. There is a linear correlation between the input voltage and the converted value (ADC). The STM32F401RE ADC range is $0-3.3V$. The ADC has a 12-bit resolution, so an input of $3.3V$ corresponds to a value of $2^{12}=4096$, and $0V$ corresponds to a value of~$0$. Details about the sensors and software used can be found in~\cite{web_CCD}.
  
\subsection{Signal processing with R}\label{subsec:processing}
\begin{figure}
      \centering
      \begin{subfigure}[b]{0.6\textwidth}
          \centering
          \includegraphics[width=\textwidth]{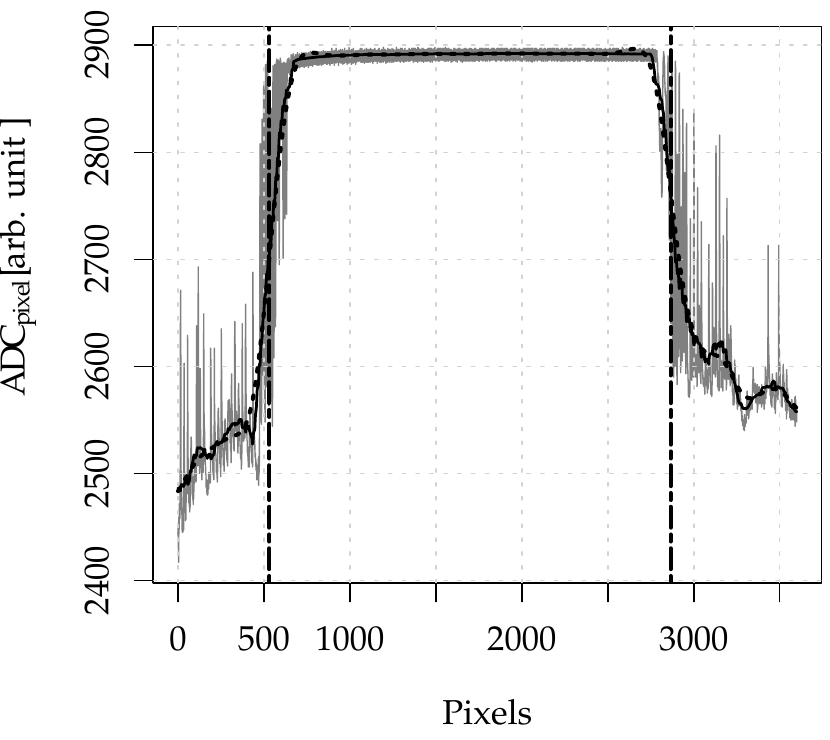}
           \caption[]%
           {{\small}}
           \label{fig:signal9_990mm_1ch}
      \end{subfigure}
      \begin{subfigure}[b]{0.6\textwidth}
          \centering
          \vskip\baselineskip
          \includegraphics[width=\textwidth]{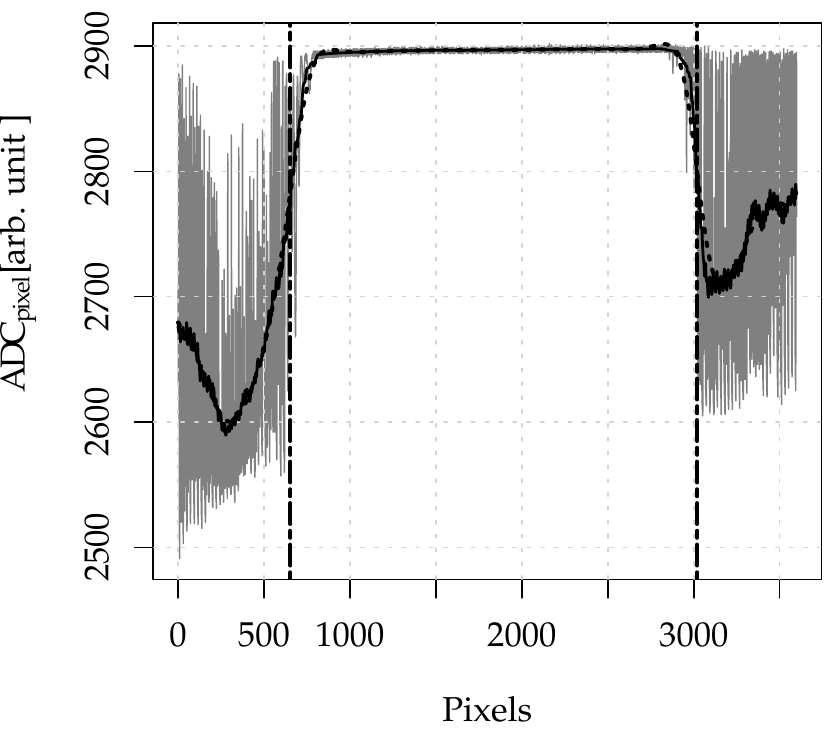}
          \caption[]%
          {{\small}}
          \label{fig:signal9_990mm_2ch}
      \end{subfigure}
      \caption[Signals depicted for both CCDs]
      {\small The output signals from the CCDs through the analogue-to-digital converters (ADCs) of the Nucleo microcontroller boards. Signals depicted (solid grey lines) for both CCDs: the first channel $\#1$ (a) and the second $\#2$ (b) for a measurement of a steel rod of $\phi=9.990$ mm. Solid black lines are filtered signals (with an average rectangle moving window of 80 pixels), and dotted lines are fitted b-splines (with degrees of freedom of 27). The two-dashed vertical lines are the cutoff corresponding to the object's shadow detected with width $P$ $[$Pixels$]$}
      \label{fig:ccd_signals}
  \end{figure}

\textsc{R} is a programming language and environment for statistical computing and graphics. Developed at Bell Laboratories, R is influential in statistics and image processing and is highly extensible via packages~\cite{R_project}.\\ 
An open-source \textsc{R} algorithm was developed to process CCD signals and to evaluate specimen dimensions. The code is freely available on GitHub~\cite{Kucharski2022_Rcode_neckle}. Open-source C programmes were used to control Nucleo boards and capture CCD data~\cite{web_CCD}. The programmes were run parallel for each board using simple Unix Bash shell commands. The signals were stored as *.dat files for further processing. For signal denoising, the moving average filter was applied (see, e.g., figs.~\ref{fig:signal9_990mm_1ch}, \ref{fig:signal9_990mm_2ch}). Next, for signal fitting, the b-spline was implemented using the \textsc{R} \textsf{stats} package with defined degrees of freedom~\cite{DeBoor1972, Perperoglou2019, Quak_2016}. The detection of the edge of the signal is based on the first derivative of the fitted signal. In this way, the width of the shadow is determined as the parameter $P$ $[$Pixels$]$. The programme steps are described in detail in appendix~\ref{sec:app_R}. The b-spline is more described in the appendix.\\
Signal processing is simple, fast, and reliable. The computation time for the $100$ files took less than $30$ s on the Unix machine based on the Intel Core i5-5250U $1.6$ GHz (without parallel computing, Geekbench score is 5094). The electronic noise on the graphs~\ref{fig:signal9_990mm_1ch} and \ref{fig:signal9_990mm_2ch} negligibly influences the precision of measurement, estimated as the standard deviation of $\overline P$ (average $P$) under $100$ repeated detections (with speed one detection per $10$ seconds) at the level of $2\sigma<5$ pixels.
\subsection{Calibration}\label{subsec:calibration}
An optical instrument needs to be calibrated. Steel rod standards with a minor diameter step of $10$ $\mu m$ and tolerance $1$ $\mu m$ were used to evaluate the resolution of the system. 
Each rod size was detected $100$ times with a detection speed of one per $10$ seconds for the average shadow width ($\overline{P}$ $[$Pixels$]$) and standard deviation ($2\sigma$) estimations. The linear fitting-based graphical method was used to find the mathematical relation between the object's size ($\phi$) and the optical output of the system (shadow width $P$). It is evident that for $\phi=0$, the shadow width (theoretically) should be equal to $0$). The fitting with an intercept equal to $0$ was checked, and the lower precision appeared than the non-zero intercept fitting, mainly affecting the fitting in the diameter range $\phi=10$ mm (for a comparison example, see the appendix). The reason comes from the detector noise as well. The estimated maths functions better represent the optical output with a free non-zero intercept than equal $0$, and the conditions were also kept because of the tested sample sizes $\phi=7-12$ mm. Moreover, a slight difference in the calibration equation may affect only the estimated diameter values of the sample, while the profile of the monitored sample diameter will remain the same.   
 
\begin{figure}
      \centering
      \begin{subfigure}[b]{0.65\textwidth}
          \centering
          \includegraphics[width=\textwidth]{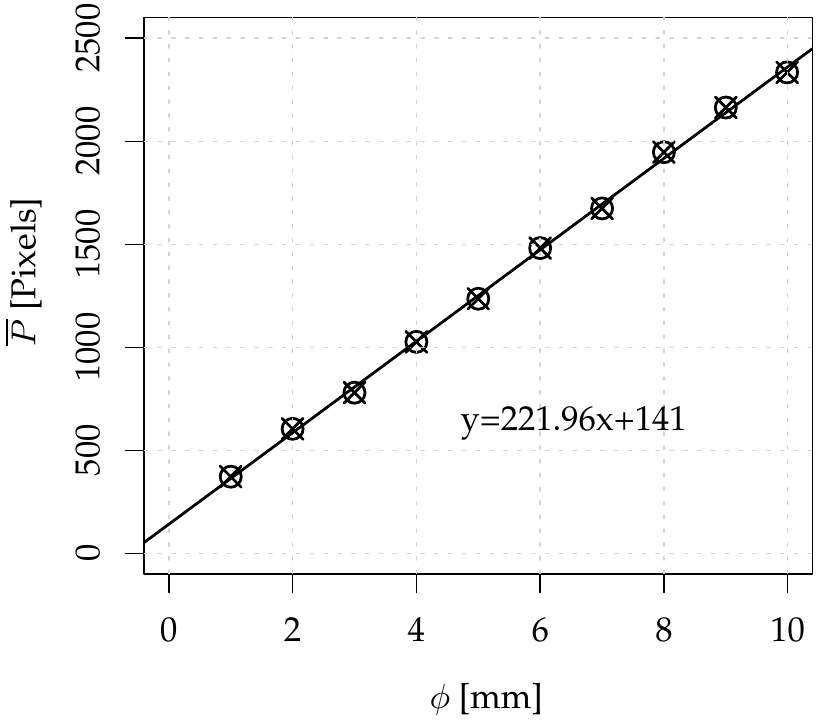}
           \caption[]%
           {{\small}}
           \label{fig:calibration_graph_1ch}
      \end{subfigure}
      \hfill
      \begin{subfigure}[b]{0.65\textwidth}
          \centering
          \vskip\baselineskip
         \includegraphics[width=\textwidth]{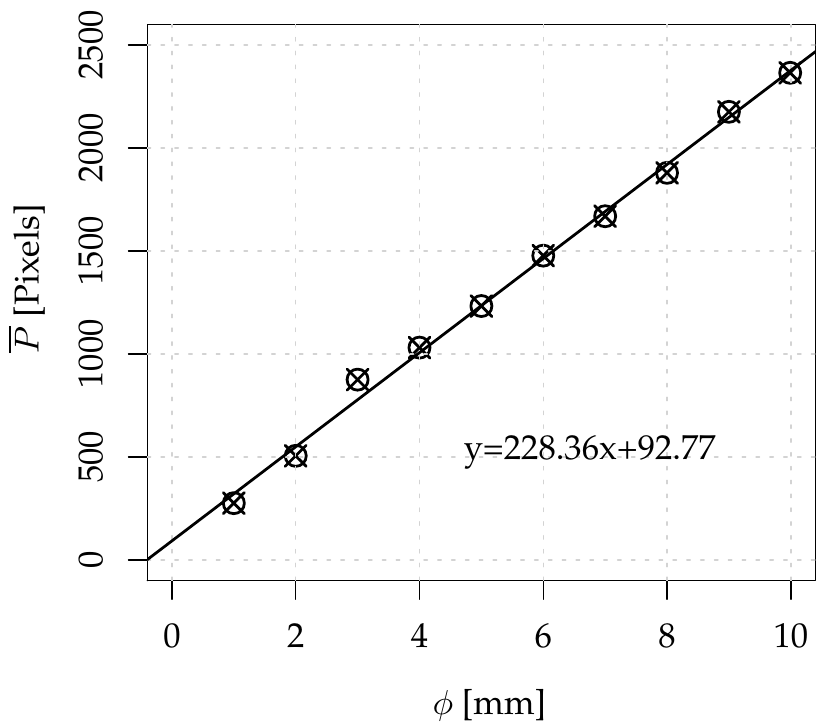}
         \caption[]%
          {{\small}}
          \label{fig:calibration_graph_2ch}
      \end{subfigure}
      \caption[Calibration graphs]
      {\small Calibration curves (linear regression $y=ax+b=\overline{P}=a\phi+b$) for: (a) -- channel $\#1$ with fitted parameters: $a=221.96$, $b=141$, std. errors $\Delta a=2.47$, $\Delta b=15.36$, $p<0.001$; (b) -- channel $\#2$ with fitted parameters: $a=228.36$, $b=92.77$, std. errors $\Delta a=5.12$, $\Delta b=31.77$, $p<0.001$}
      \label{fig:calibration_graphs}
  \end{figure}

The calibration graphs of the two setup channels are depicted in fig.~\ref{fig:calibration_graphs}. The $1$ $mm$ of the diameter step was chosen for the clarity of the graphs. The linear functions are used to monitor the sample diameter during tensile tests. An example of a sub-millimetre optical response is shown in table~\ref{tab:submm}. 

\begin{table}
\centering
\caption{Example of measured calibration steel rods with $10$ $\mu m$ diameter differences. $\#1$ -- channel 1, $\#2$ -- channel 2, $\phi$ -- diameters in $[$mm$]$, $\overline{P}$ -- average shadow's width in $[$Pixels$]$,  $2\sigma$ -- $2 \times$standard deviation in $[$Pixels$]$.}
\label{tab:submm}
\begin{tabular}{ccccccc}
\hline\hline
\multicolumn{5}{c}{$\overline{P}\pm 2\sigma$ $[$Pixels$]$} \\ \hline
$\phi$ $[$mm$]$ & $9.99$ & $9.98$ & $8.01$ & $8.00$ & \\
$\#1$ & $2334.05\pm 1.60$ & $2330.10\pm 1.20$ & $1966.61\pm 1.61$ & $1946.58\pm 1.22$ &  \\
$\#2$ & $2364.59\pm 1.18$ & $2352.96\pm 1.51$ & $1896.39\pm 4.70$ & $1879.00\pm 1.28$ \\ \hline 
\end{tabular}
\begin{tabular}{ccc}
\hline\hline
\multicolumn{3}{c}{$\overline{P}\pm 2\sigma$ $[$Pixels$]$} \\ \hline
$\phi$ $[$mm$]$ & $7.01$ & $7.00$ \\
$\#1$ & $1687.00\pm 0.70$ & $1673.88\pm 1.72$ \\
$\#2$ & $1694.48\pm 1.47$ & $1668.59\pm 1.18$
\end{tabular}

\end{table}

\subsection{Samples and experimental workflow}\label{subsec:workflow}
\begin{figure}
\centering	
\begin{minipage}[t]{.45\textwidth}
\centering
  \includegraphics[width=1\textwidth]{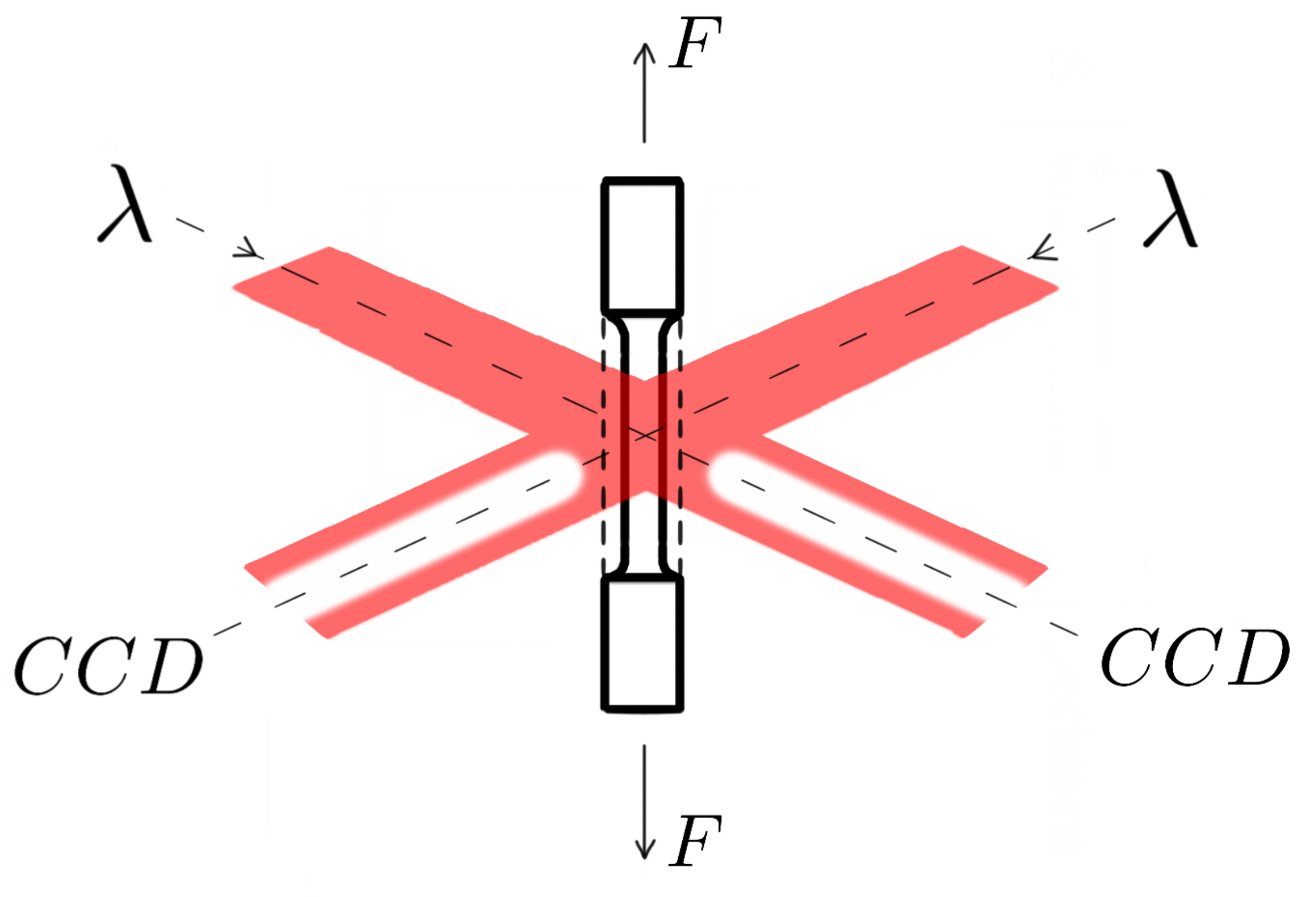}
\captionof{figure}{The experimental scheme: $\lambda$'s - laser light sources, $CCD$s - array detectors, $F$ - elongation force}
  \label{fig:workflow}

\end{minipage}%
\hfill
\centering
\begin{minipage}[t]{.45\textwidth}
\centering
  \includegraphics[width=\textwidth]{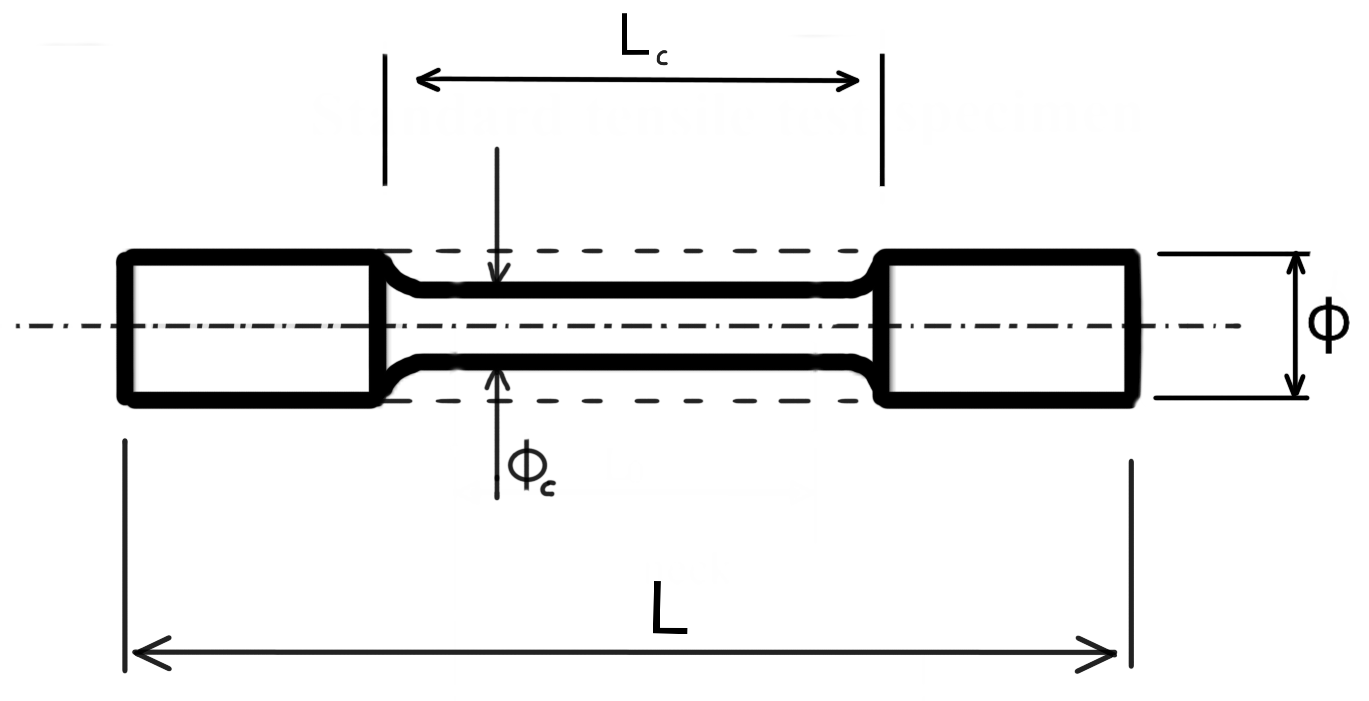}
 \captionof{figure}{Sample dimensions: $\Phi\approx  12$ mm, $\Phi_c \approx  10$ mm, $L\approx 20$ cm, $L_c\approx 3$ cm}
  \label{fig:sample_size}

\end{minipage}
\end{figure}

 The experiment scheme is shown in fig.~\ref{fig:workflow}. PET (polyethylene terephthalate), PVC (polyvinyl chloride) and PVDF (polyvinylidene fluoride) rods with diameter $\phi\approx 12$ mm were used as samples. Pure and precut rods (with approximately $\Delta \phi = 2$ mm initiation of the neck region) were tested (see fig.~\ref{fig:sample_size}).
 
  Engineering stress $\sigma$ is the applied load $F$ measured by the machine sensor, divided by the original cross-sectional area of material $A_{0}$, calculated from the initial diameter of a sample (see eq. \ref{eq:stress}).
  
\begin{equation}
	\sigma=\frac{F}{A_0}=\frac{F}{\pi r^2},
	\label{eq:stress}
\end{equation}
where: $F$ -- applied force, $A_{0}$ -- original cross-sectional area, $r$ -- sample radius.\\
 The strain is the amount of sample deformation due to an applied force. The strain ($\epsilon$) is dimensionless, defined as the fractional change in length (see eq.\ref{eq:strain}).
 
\begin{equation}
	\epsilon=\frac{\Delta L}{L_0},
	\label{eq:strain}
\end{equation}
where: $\Delta L$ -- lenght change, $L_{0}$ -- initial length.\\
In the results presented, $L_{0}$ corresponded to the initial position of the upper mounting bracket of the machine without load and $\Delta L$ was the displacement of the mounting bracket detected by the machine during the test.

\begin{figure}
      \centering
      \begin{subfigure}[b]{0.475\textwidth}
          \centering
          \resizebox{1\textwidth}{!}{%
          \includegraphics[width=0.6\textwidth]{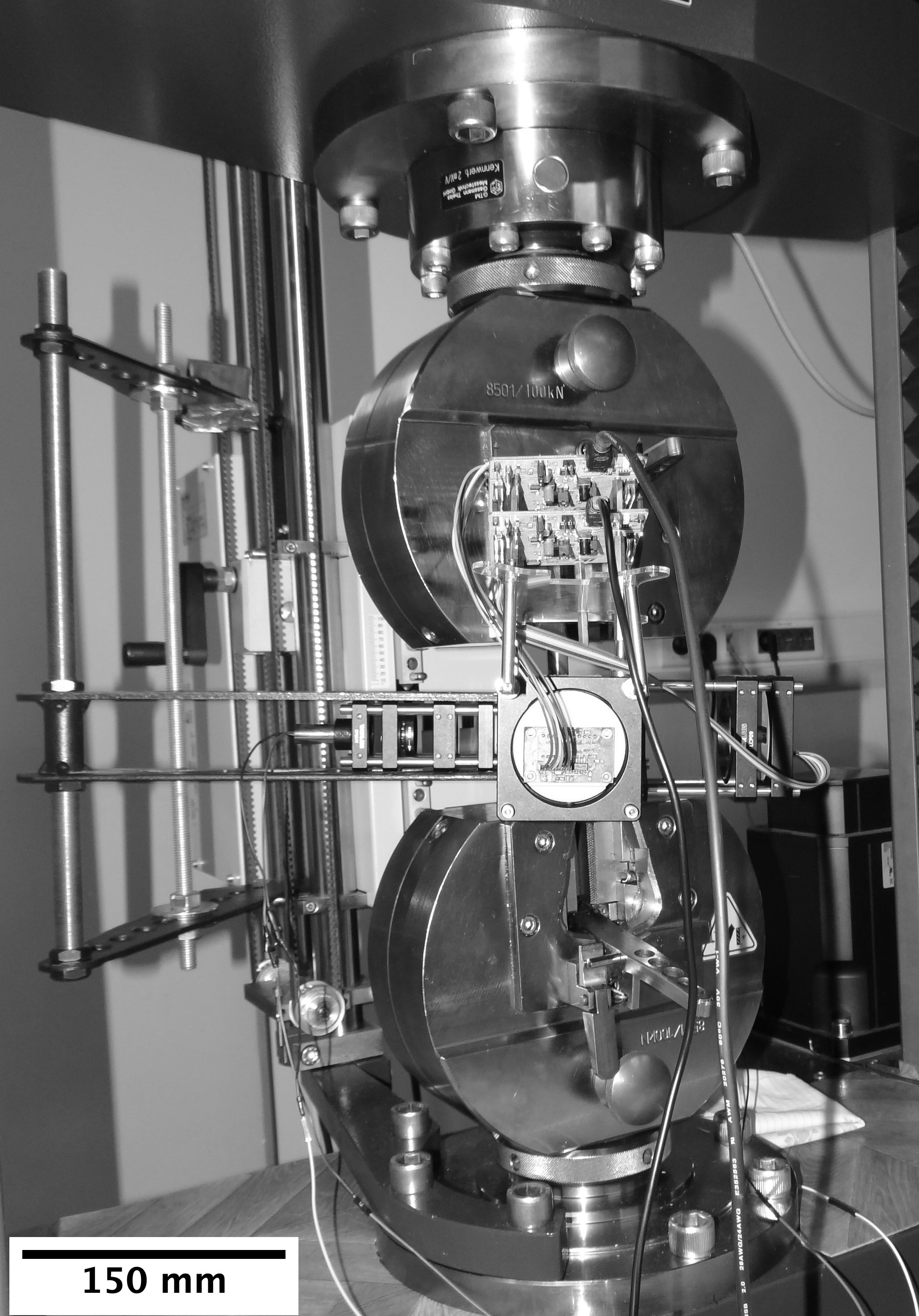} 
           }%
           \caption[]%
           {{\small}}
           \label{fig:projector_on_machine_3}
      \end{subfigure}
      \hfill
      \begin{subfigure}[b]{0.4\textwidth}
          \centering
          \resizebox{1\textwidth}{!}{%
          \includegraphics[width=0.2\textwidth]{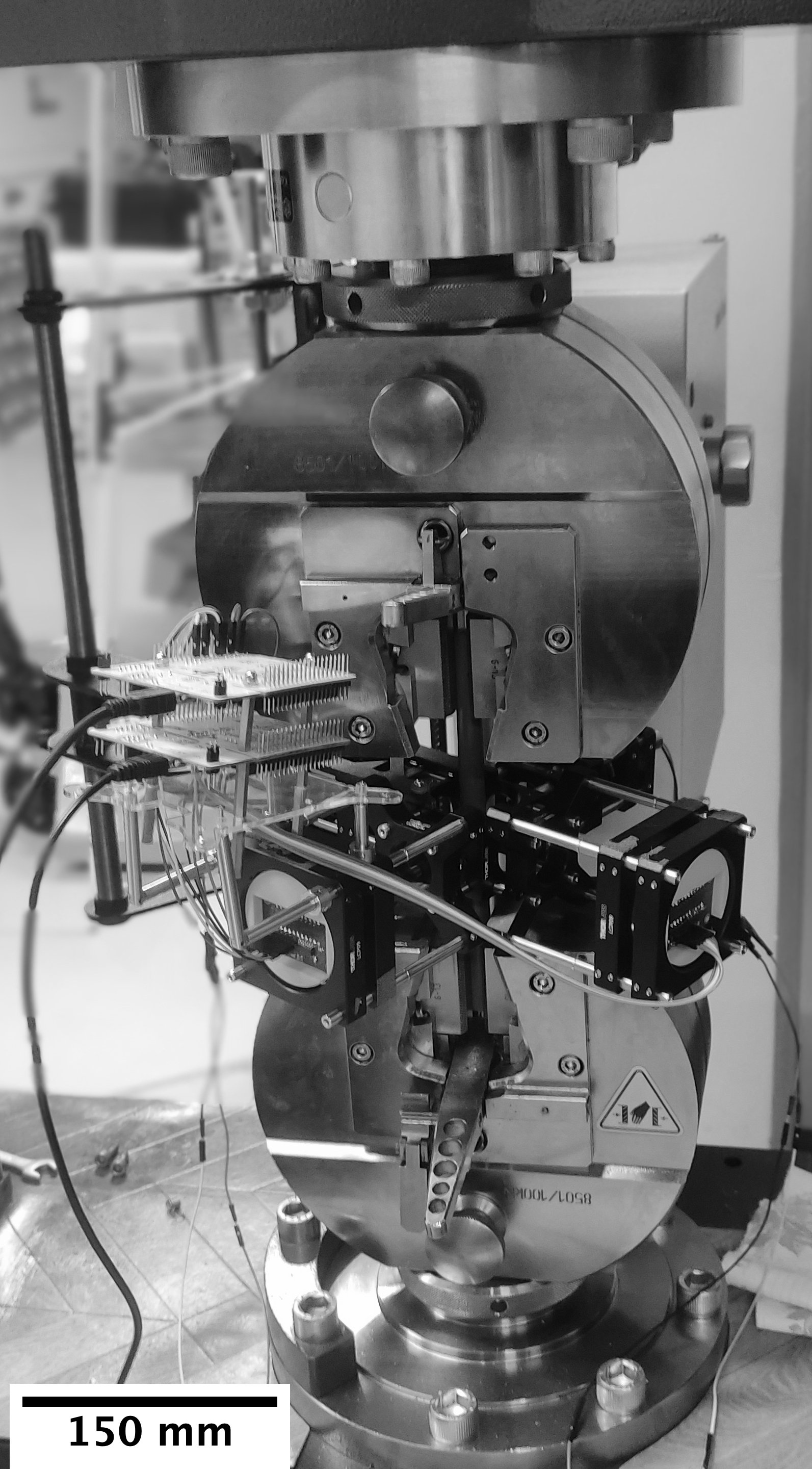}
          }%
         \caption[]%
          {{\small}}
          \label{fig:projector_on_machine_1}
      \end{subfigure}
      \caption[The projector on machine]
      {\small The projector on the tensile machine: (a) front view, (b) side view}
      \label{fig:projector_on_machine}
  \end{figure}

\noindent The Zwick Z100 machine was used for tensile tests with the optical projector attached to the moving traverse beam by a specially designed stiff arm (see fig.~\ref{fig:projector_on_machine}). The maximum load is $10$ $kN$. The fixing point is in the centre of the projector (the centre cube). During data acquisition, the optical setup on the column was moved at half the elongation speed by the transmission belt. For the first tests, it was $2.5$ mm/min. This speed kept the laser beams close to the possible necking region. The reduced speed avoids the second necking initiation in the sample and is chosen for the setup performance tests. It is worth emphasising that the optical setup was attached to the machine, replacing an extensometer, and preventing in this way, the ability to simultaneously monitor the elongation of the sample in the area of the optical projector detection. The approach to the correlation between displacement and necking is shown, and a different statistical approach will be proposed in the future to analyse the mechanical properties obtained with an extensometer. 

\section{Results and discussion}\label{sec:results}  
\begin{figure}
\begin{subfigure}{.495\linewidth}
\centering
          \includegraphics[width=\textwidth]{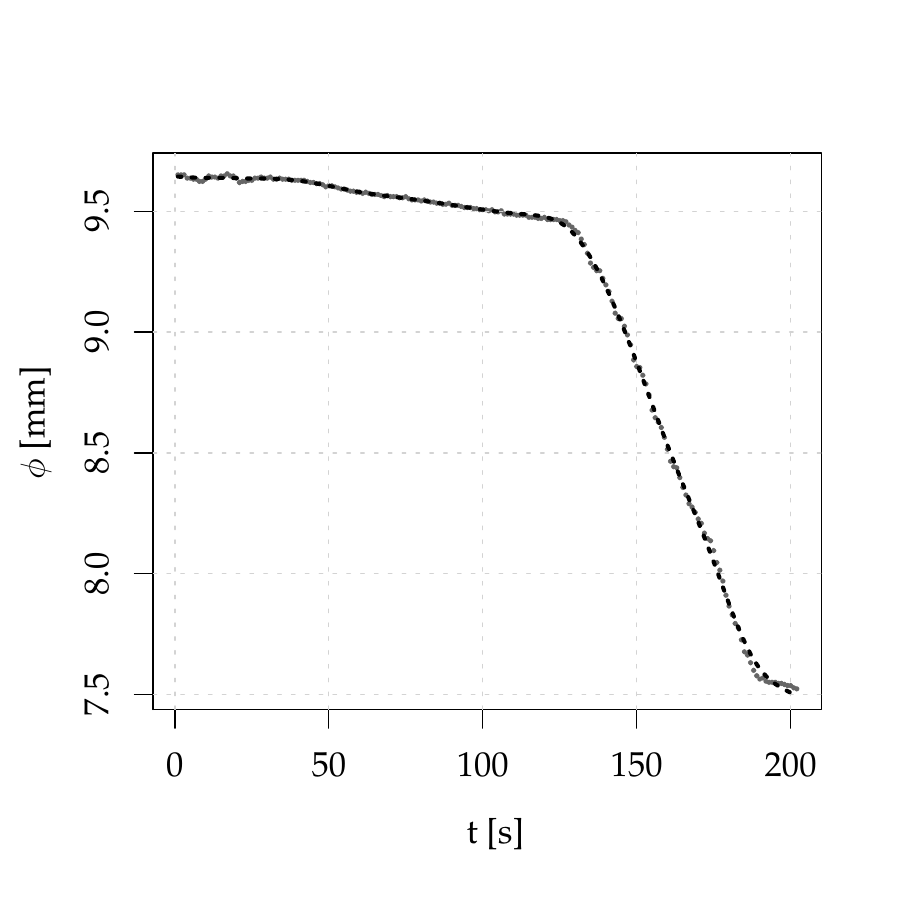}
           \caption[]%
           {{\small}}
           \label{fig:pet_neckled_graph_1ch}
\end{subfigure}%
\hfil
\begin{subfigure}{.495\linewidth}

\centering
          \includegraphics[width=\textwidth]{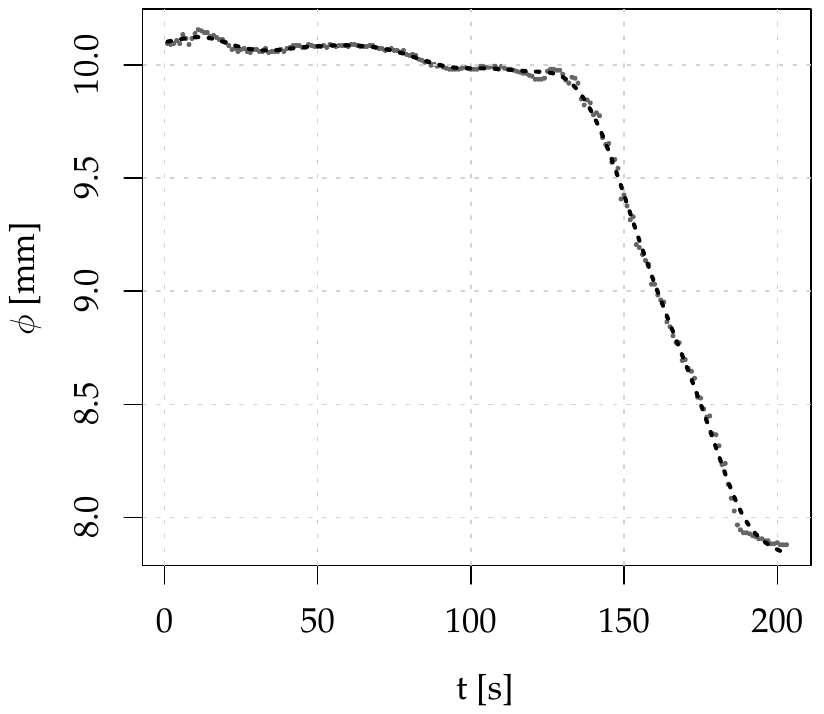}
           \caption[]%
           {{\small}}
           \label{fig:pet_neckled_graph_2ch}
\end{subfigure}
\begin{center}
\begin{subfigure}{.65\linewidth}
\vspace\baselineskip
\includegraphics[width=\textwidth]{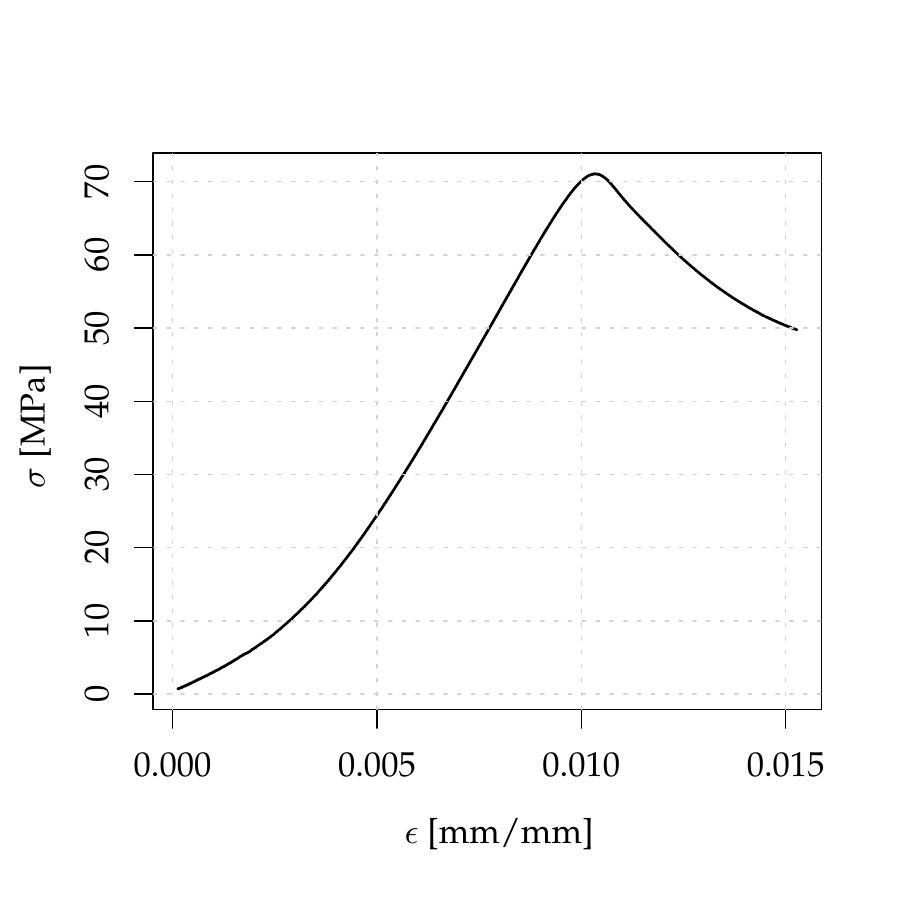}
           \caption[]%
           {{\small}}
           \label{fig:pet_neckled_stress}          
\end{subfigure}
\end{center}
\caption[PET_neckled_results]
      {\small Example results for a precut PET polymer rod: (a) - diameter change registered by channel $\#1$, (b) - diameter change registered by channel $\#2$, (c) - stress-strain curve obtained by the tensile machine. The initial force $F_{max}\approx 5.2$ $kN$. The grey points correspond to registered data; the dashed line corresponds to spline fitting}
      \label{fig:pet_neckled}           
\end{figure}

\begin{figure}
\begin{subfigure}{.495\linewidth}
\centering
          \includegraphics[width=\textwidth]{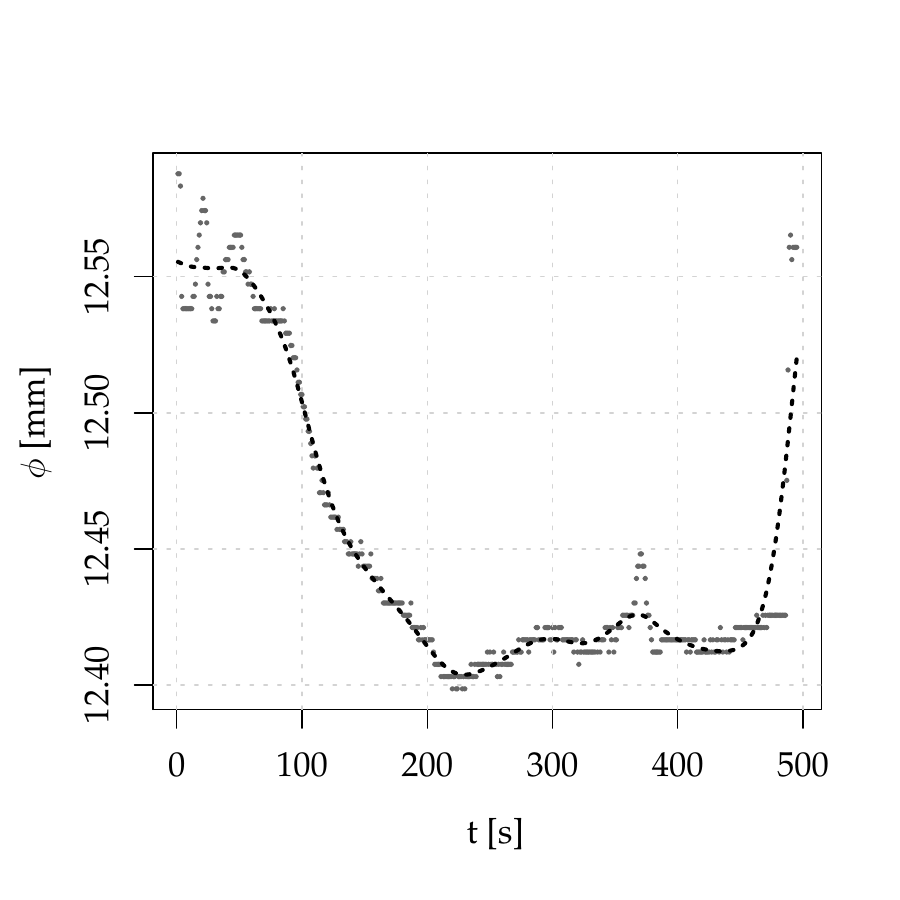}
           \caption[]%
           {{\small}}
           \label{fig:pet_pure_graph_1ch}
\end{subfigure}%
\hfill
\begin{subfigure}{.495\linewidth}
\centering
          \includegraphics[width=\textwidth]{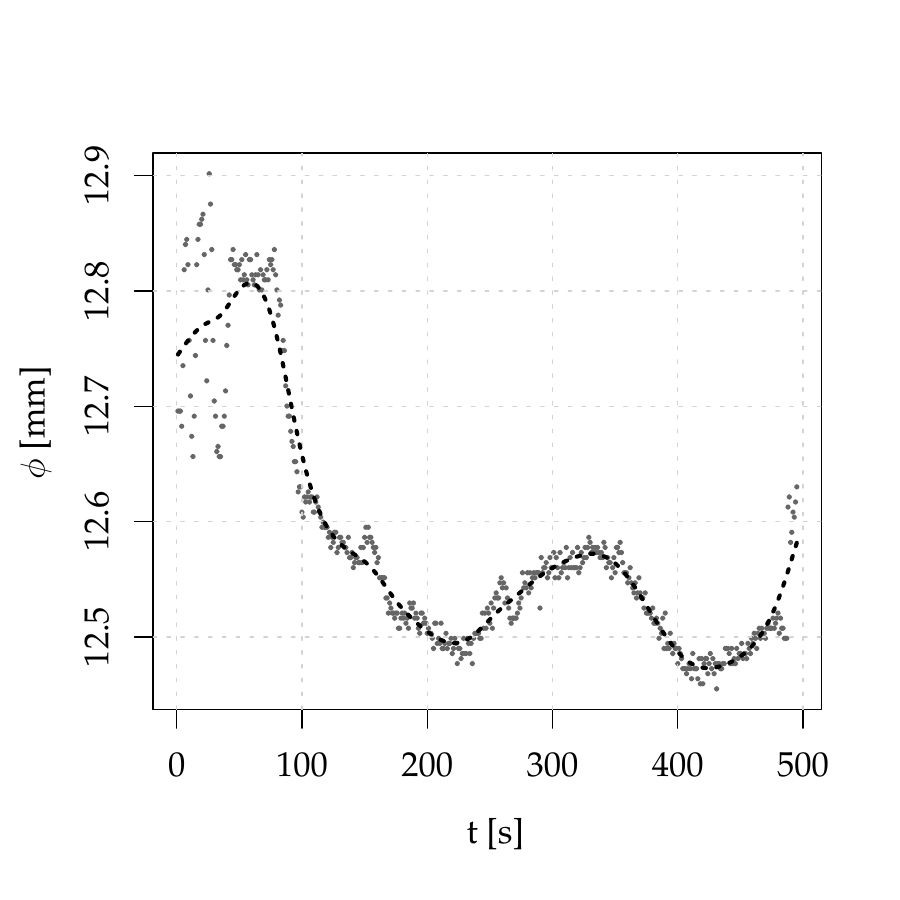}
           \caption[]%
           {{\small}}
           \label{fig:pet_pure_graph_2ch}
\end{subfigure}
\begin{center}
\begin{subfigure}{0.65\linewidth}
\vspace\baselineskip
          \includegraphics[width=\textwidth]{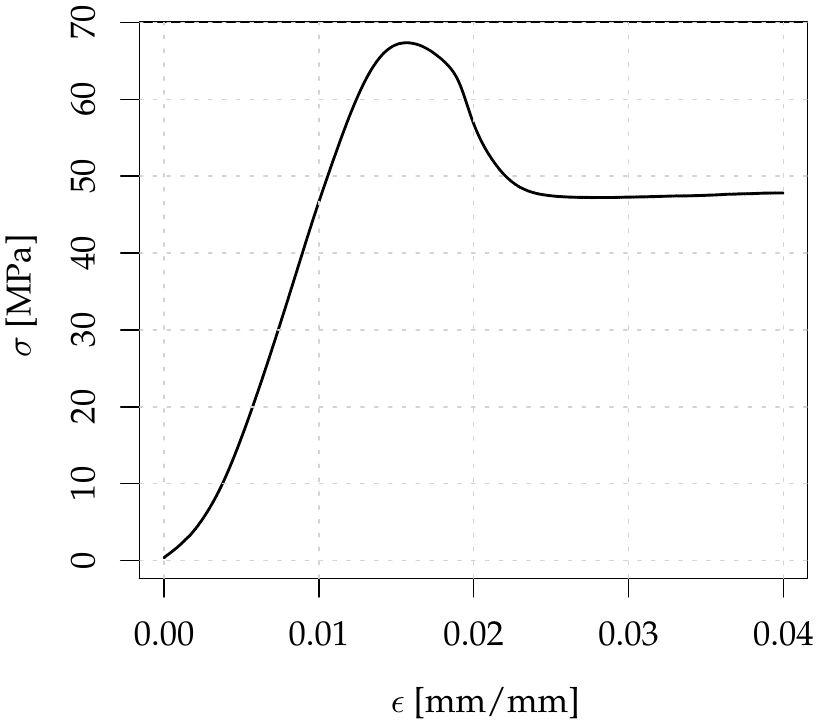}
           \caption[]%
           {{\small}}
           \label{fig:pet_pure_stress}
\end{subfigure}
\end{center}
      \caption[PET_pure]
      {\small Example results for a solid PET polymer rod: (a) - diameter change registered by channel $\#1$, (b) - diameter change registered by channel $\#2$, (c) - stress-strain curve obtained by the tensile machine. The initial force $F_{max}\approx 8.5$ $kN$. The grey points correspond to registered data; the dashed line corresponds to spline fitting}
      \label{fig:pet_pure_results}
  \end{figure}

Eight measurements in total were analysed for the first setup tests. The results proved the remarkable reliability of the constructed setup. The example results shown in the paper proved the complexity, symmetry, and non-symmetry of the polymer's mechanical behaviour. This corresponds to the two measurement directions for the sample's diameters (to the symmetric or non-symmetric response of the channels). In fig.~\ref{fig:pet_neckled}, the measurement results for the precut PET rod are presented. The sample diameter profiles in both transverse directions are similar and correspond to the detection in the neck region, where the sample finally breaks. The b-splines with degrees of freedom of $15$ were applied to smooth the profiles (for an additional description, see the appendix). In fig.~\ref{fig:pet_neckled_stress}, the maximum of the stress-strain curve named the ''necking point'' corresponds to the fast radius reduction of the sample after approximately $130$ seconds of elongation. The initial force $F_{max}$ corresponding to the maximum stress was approximately $5.2$ $kN$. The data were stored every second for maximum precision. No time synchronisation between the optics and the machine was prepared for the first tests. Research with a specially prepared programme will begin presently.\\
The results obtained show some asymmetry for the same polymer PET without pre-necking preparation (precut) (see figs.~\ref{fig:pet_pure_graph_1ch} and~\ref{fig:pet_pure_graph_2ch}). The initial force corresponding to the maximum stress was $F_{max}\approx 8.5$ $kN$. At the end of the test, the diameter growth is presented. This group of points was detected after the sample cracked. The rest of the sample still created the shadow. This is related to the necking region shift initiated off the optical axis and the sample shift after the crack. The effect will be further investigated with time synchronisation.\\
Moreover, some fluctuations are observed during the first $50$ seconds of the process. The effect corresponds to the machine pre-set procedure. Before the direct procedure, the machine moved up and down (with a force of $0N - 50N - 0N$) to avoid slippage of the polymer by increasing the hold of the clamps by biting into the material. This affects probe radius oscillation and is observed in the much smaller range for the probe with the necking precut (see figs.~\ref{fig:pet_neckled_graph_1ch} and~\ref{fig:pet_neckled_graph_2ch}). This procedure does not significantly affect the sample's mechanical properties in this range. It is not detected even by the machine sensors with such a small stress range. The effect might be considered for the subsequent modelling properties using the proposed optical measurements.
 
\begin{figure}
\begin{subfigure}[b]{.48\linewidth}
\centering
         \includegraphics[width=\textwidth]{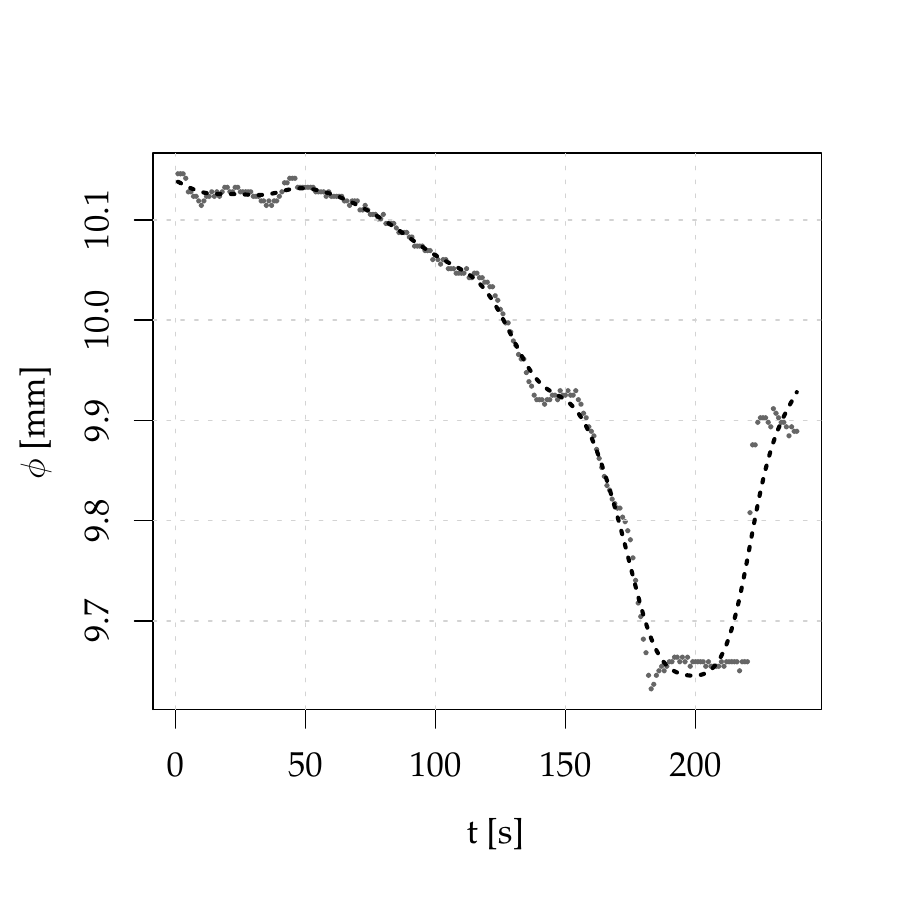}
           \caption[]%
           {{\small}}
           \label{fig:pvdf_neckled_graph_1ch}
\end{subfigure}%
\hfill
\begin{subfigure}[b]{.48\linewidth}
\centering
          \includegraphics[width=\textwidth]{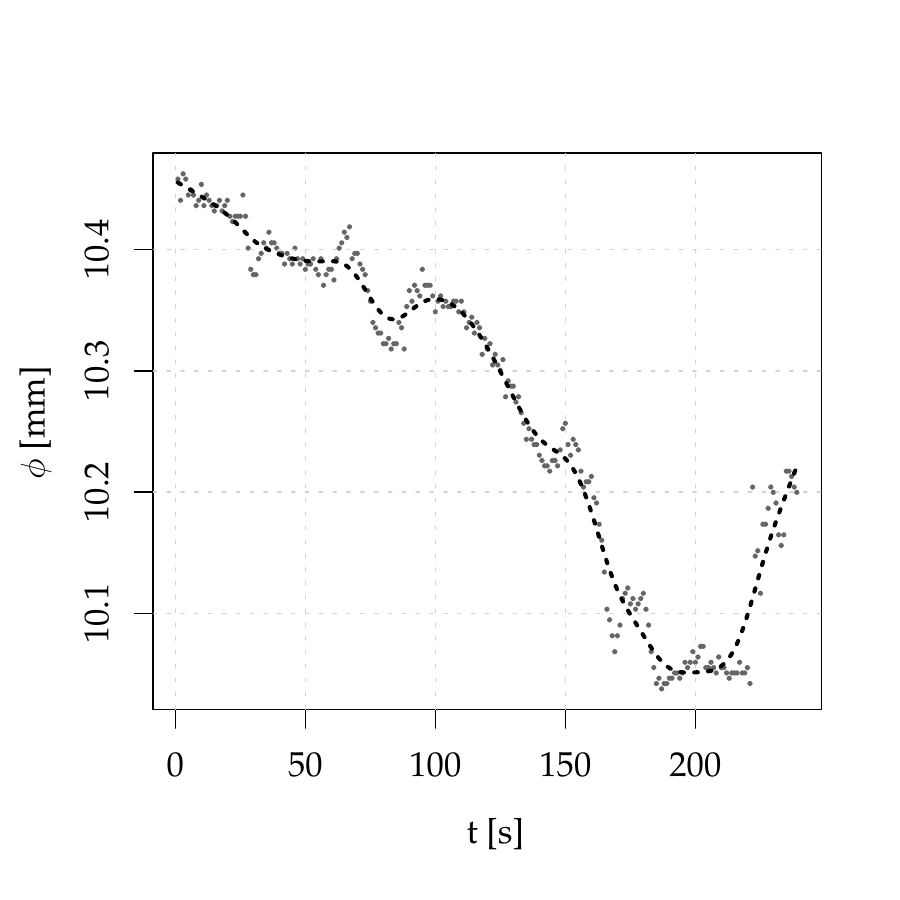}
           \caption[]%
           {{\small}}
           \label{fig:pvdf_neckled_graph_2ch}
\end{subfigure}

\begin{subfigure}[b]{.48\linewidth}
\vspace\baselineskip
\centering
          \includegraphics[width=\textwidth]{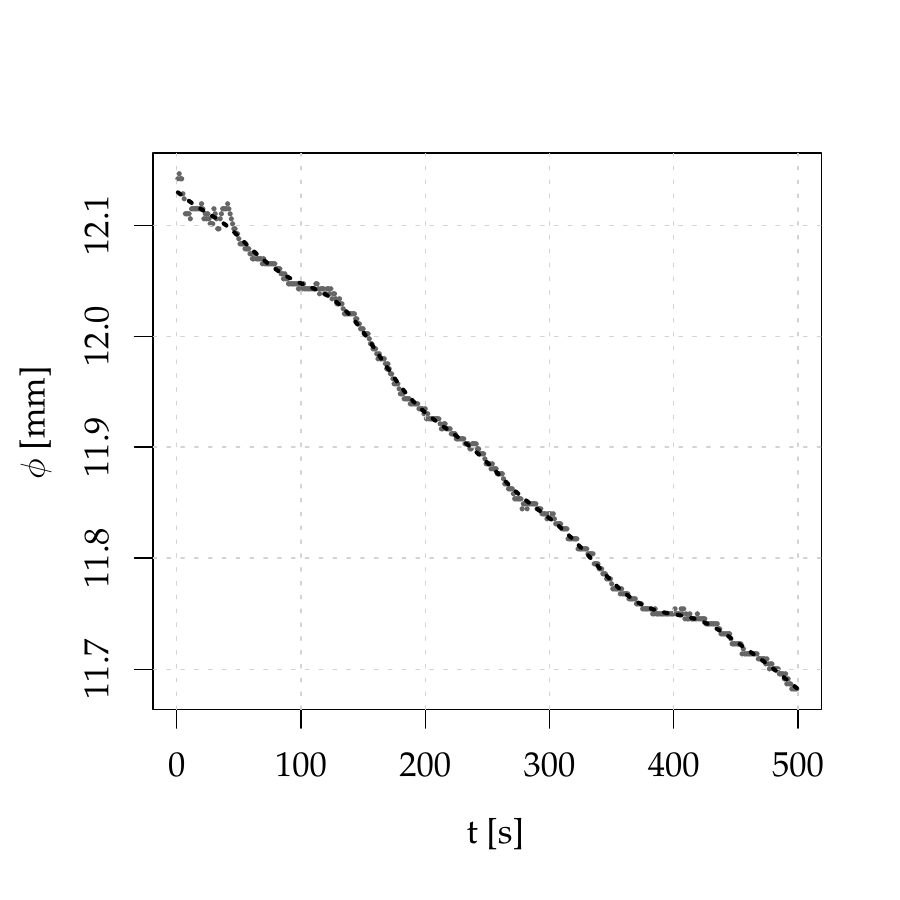}
           \caption[]%
           {{\small}}
           \label{fig:pvdf_pure_graph_1ch}
\end{subfigure}
\hfill
\begin{subfigure}[b]{.48\linewidth}
\centering
          \includegraphics[width=\textwidth]{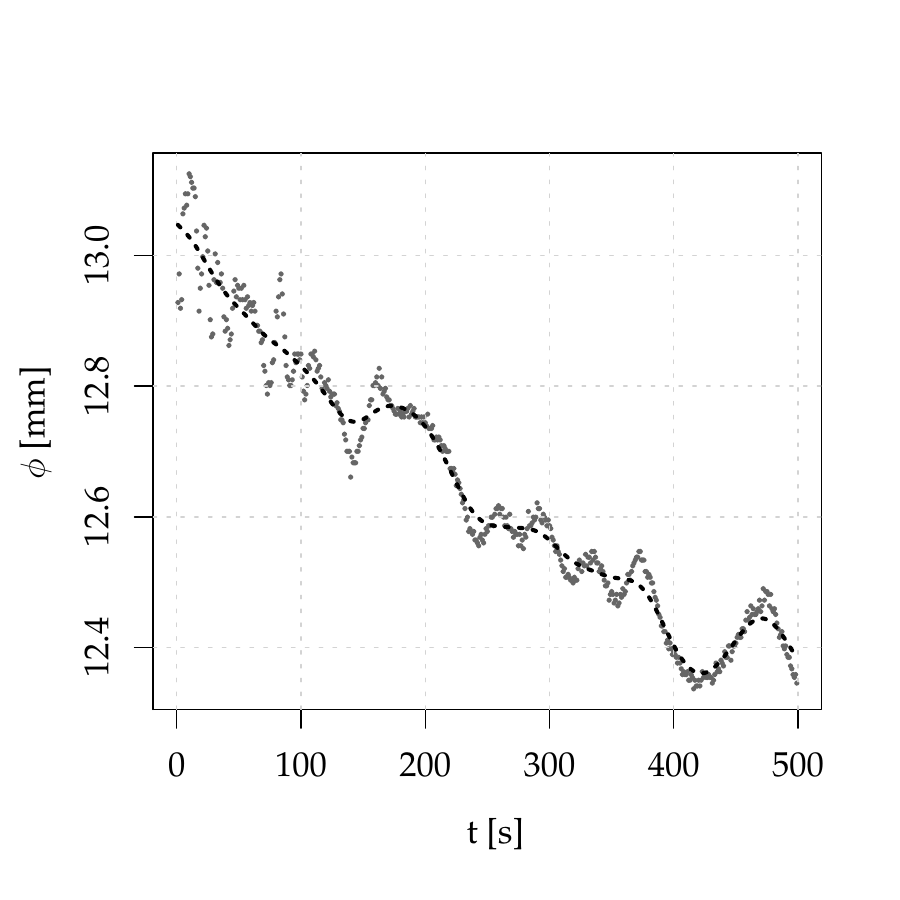}
           \caption[]%
           {{\small}}
           \label{fig:pvdf_pure_graph_2ch}
\end{subfigure}

\caption[PVDF results]
      {\small Example results for PVDF polymer rod: (a) - diameter change registered by channel $\#1$ for a precut sample, (b) - diameter change registered by channel $\#2$ for a precut sample. The initial force $F_{max}\approx 4.6$ $kN$, (c) - diameter change registered by channel $\#1$ for a solid sample, (d) - diameter change registered by channel $\#2$ for a solid sample. The initial force $F_{max}\approx 5.4$ $kN$}
      \label{fig:pvdf_results}           
\end{figure}

 \noindent An apparent difference between stress-strain curves shows sharper necking initiation in the precut sample (see fig.~\ref{fig:pet_neckled_stress}) than in the solid one (see fig.~\ref{fig:pet_pure_stress}).\\ 
An interesting effect was observed during the PVDF sample measurements. The optics detected the initiation of the necking in the precut sample (see Figs.~\ref{fig:pvdf_neckled_graph_1ch} and~\ref{fig:pvdf_neckled_graph_2ch}) and the absence of the necking process in the solid polymer rod (see fig.~\ref{fig:pvdf_pure_graph_1ch} and~\ref{fig:pvdf_pure_graph_2ch}). The effect might be related to the homogeneity of the samples and the repeatability of the measurements. This will be considered for further investigation and metrology evaluation of the setup. The growth of the sample diameter at the end of the test for the precut PVDF is again related to the points detected after the sample breaks and shifts. The initial force reached $F_{max}\approx 4.6$ $kN$ for the precut sample and $F_{max}\approx 5.4$ $kN$ for the solid sample, respectively. 
 

\begin{figure}
\begin{subfigure}[b]{.48\linewidth}
\centering
          \includegraphics[width=\textwidth]{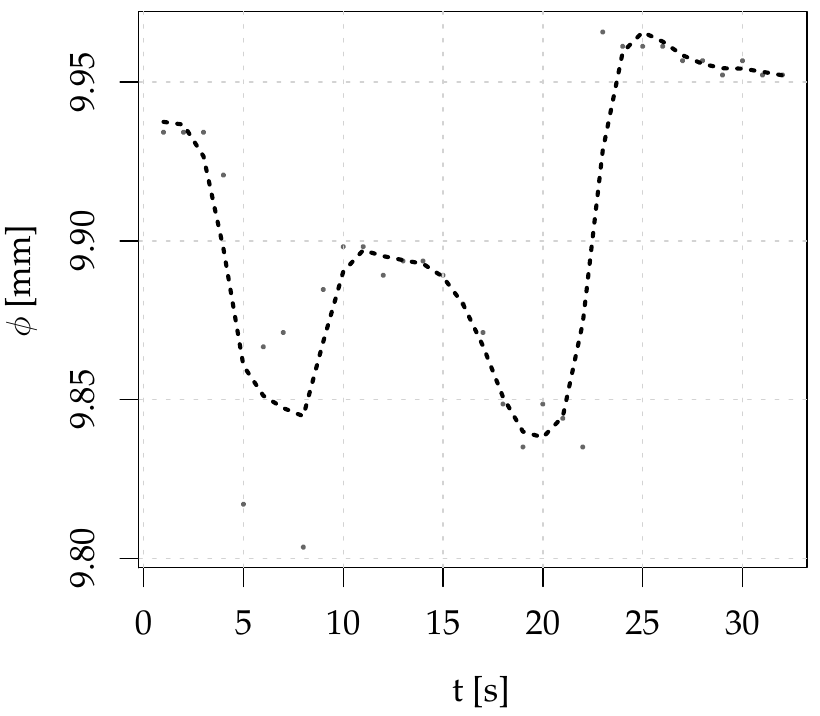}
           \caption[]%
           {{\small}}
           \label{fig:pvc_neckled_graph_1ch}
\end{subfigure}%
\hfill
\begin{subfigure}[b]{.48\linewidth}
\centering
          \includegraphics[width=\textwidth]{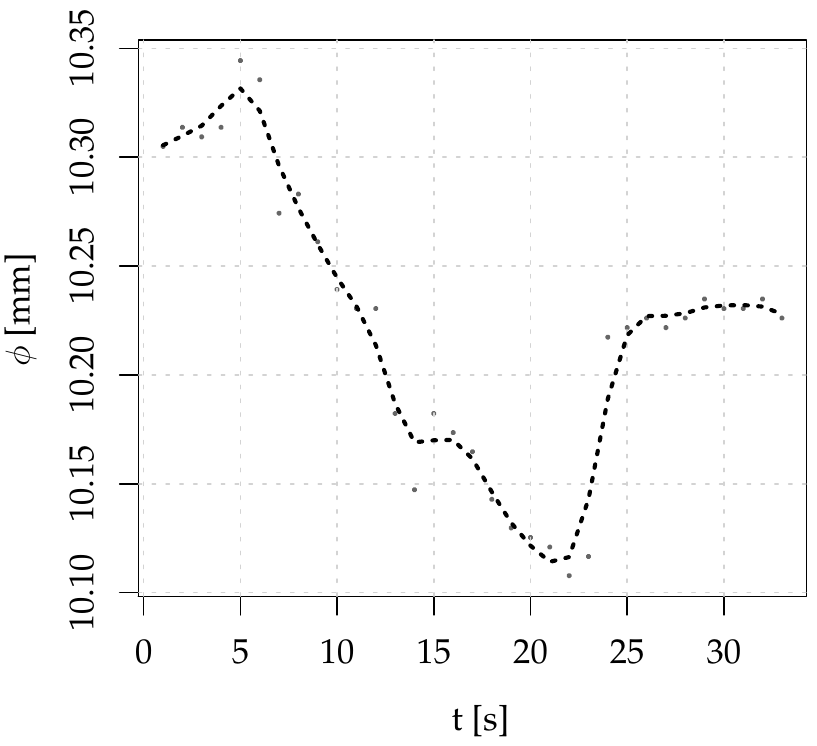}
           \caption[]%
           {{\small}}
           \label{fig:pvc_neckled_graph_2ch}
\end{subfigure}
\begin{subfigure}[b]{.48\linewidth}
\vspace\baselineskip
\centering
          \includegraphics[width=\textwidth]{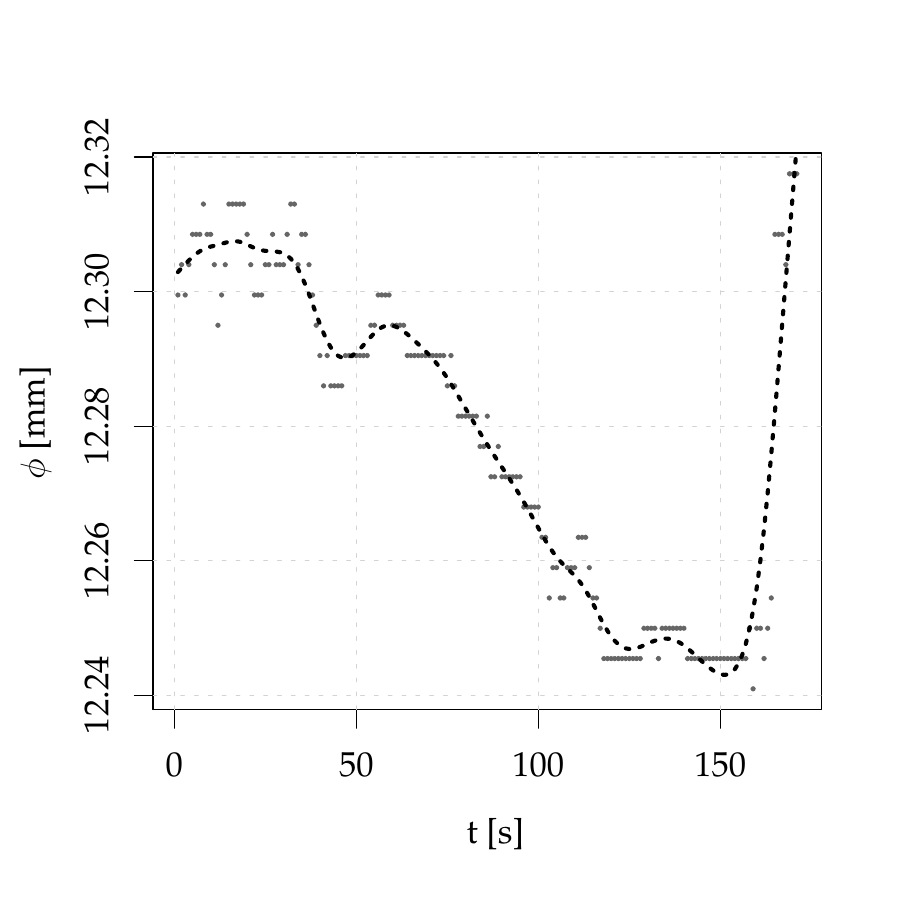}
           \caption[]%
           {{\small}}
           \label{fig:pvc_pure_graph_1ch}
\end{subfigure}
\hfill
\begin{subfigure}[b]{.48\linewidth}
\centering
          \includegraphics[width=\textwidth]{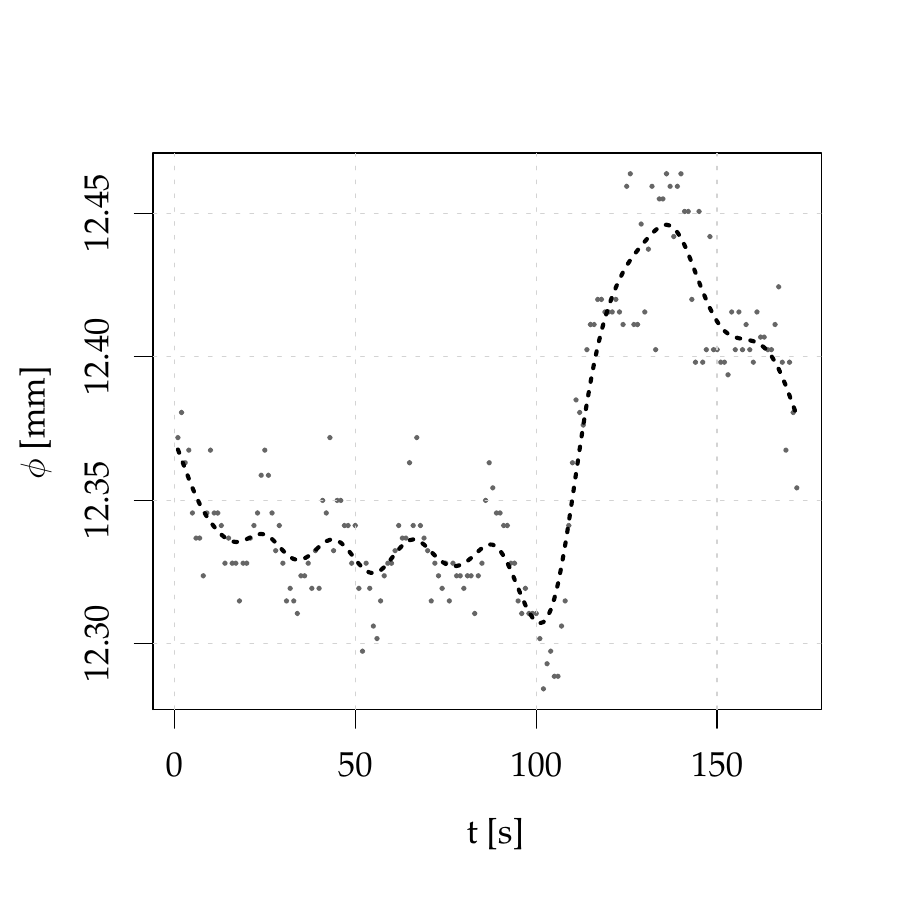}
           \caption[]%
           {{\small}}
           \label{fig:pvc_pure_graph_2ch}
\end{subfigure}

\caption[PVC results]
      {\small Example results for a PVC polymer rod: (a) - diameter change registered by channel $\#1$ for a precut sample, (b) - diameter change registered by channel $\#2$ for a precut sample. The initial force $F_{max}\approx 3.4$ $kN$, (c) - diameter change registered by channel $\#1$ for a solid sample, (d) - diameter change registered by channel $\#2$ for a solid sample. The initial force $F_{max}\approx 4.5$ $kN$}
      \label{fig:pvc_results}           
\end{figure}

Complex results were obtained for measurements of PVC samples. For the precut rod, neck initialisation was observed in both channels after approximately $20 s$ of elongation (see figs.~\ref{fig:pvc_neckled_graph_1ch} and~\ref{fig:pvc_neckled_graph_2ch}), while for the solid sample in the second channel $\# 2$ periodic oscillations and diameter growth were observed after $100 s$ of elongation, opposite to the channel $\# 1$. The last detected points for the precut sample after approximately $25 s$ and $150 s$ for the solid sample correspond to the cracked rod with the shifted position on the optical axis of the setup. The angle between the sample and the optical axis may change with cracking and gives an asymmetric response in the transverse channels of the setup. Furthermore, depending on the homogeneous structure of the samples, the responses in the two channels could be different. This proves the importance of direct bi-axial transverse investigations using the optical setup. The behaviour of the material corresponds to the complexity of the polymer structure and the response to stress. Mechanical behaviour under tensile tests may not be repeatable for the same polymer samples. The obtained results are crucial for further investigations with specially selected, more homogeneous samples, e.g. prepared with a particular form during manufacturing. The results show that the samples must be precise and repeatable precuts to detect the start of the neck to increase the measurements' repeatability.
Reducing the number of measurement points can omit the sample vibration problem (see the appendix).

As can be seen, from the results and the calibration procedure, the proposed setup provides $\mu m$ resolution for sample diameter detection during tensile tests. The proposed low-cost setup may be considered the new measurement tool for evaluating the direct mechanical properties of polymer materials. The data obtained may be used to model the material's properties numerically. The repeatability and reproducibility of measurements with the tensile machine and optical setup will begin presently. 

\section{Conclusions}\label{sec:results} 
The self-built laser projection system for symmetric transverse strain measurements of polymers in tensile or compression tests has been presented. The proposed device and measurement methodology allow for monitoring of diameter changes in two perpendicular directions during specimen deformation. The novel method, the constructed and calibrated setup, has been presented with the first polymer tests using the tensile machine. The setup provided monitor polymer diameters that can change from $12 mm$ to even $1mm$ with resolution $10$~$\mu m$.
The maximum change in the diameter of the samples presented was approximately $2mm$ before cracking a polymer rod. The response of the sample to the load may be different in the orthogonal direction even at the level of $50$ $\mu m$ (for example, see figs.~\ref{fig:pet_pure_graph_1ch} and~\ref{fig:pet_pure_graph_2ch}). For all of the results, the polymer sample's diameters decreased with elongation, which confirmed the setup's performance for monitoring on the tensile machine. The noise of the data during measurements can be reduced with the number of detected points (see the appendix). For data processing, the \textsc{R} programme was prepared. The code is freely available on the Internet.\\
The optical setup was attached to the machine, replacing an extensometer and preventing, in this way, the ability to simultaneously monitor the elongation of the sample in the exact area of optical projector detection. Following this issue, precision measurements will be conducted with well-prepared repeatable samples, separately from the devices, and a statistical approach will be proposed to analyse mechanical properties. The point is synchronisation between machine and projector and repeatable conditions (homogeneous samples, precise, repeatable geometry). In this way, the repeatability and reproducibility of the measurements could also be evaluated.\\
The results have clearly shown that the proposed optical system may be a vital tool for revising correct models of the mechanical properties of polymers under complex loads. The low-cost setup provides fast and accurate sample transversal dimension monitoring in tensile or compression tests, and further work will focus on the robustness and compactness of the device. More accurate and modern material models can be proposed based on registered data.\\
More comprehensive material research will begin with specially prepared geometric and homogeneous material samples.

Notable progress in materials engineering requires more detailed research on the mechanical properties of materials. Research on non-classical materials requires adequate experimental results. The proposed methodology can be applied for the comprehensive test of mechanical properties of non-classical and linear or nonlinear behaviour materials, which can be more relevant in material properties modelling. An additional advantage is the ability to test samples of materials with a cylindrical or flat shape. The standard tensile or compression test does not provide enough information on the mechanical properties of non-classical materials such as a broad group of polymers. Accurate registration of changes in vertical specimen dimensions is necessary to describe modern material models. The appropriate material model is the basis for all structural analyses. The proposed measurement system can be an excellent tool for registering experimental data and mathematical descriptions of the material's mechanical properties under complex load conditions. The proposed open-source data processing algorithms deliver the results' clarity and full reproducibility potential. 

\section*{Declaration of Competing Interest}

The authors declare that they have no known competing financial interests or personal relationships that could appear to influence the work reported in this document.

\section*{Acknowledgments}
The authors acknowledge Professor Szymon Wojciechowski of the Faculty of Mechanical Engineering and Management, Poznan University of Technology, for constructive criticism of the manuscript.\\
Funding: This work was supported by grant DS 061/SBAD/1547.

\appendix
\section{The signal processing details}\label{sec:app_R}
For signal processing, \textsf{pracma} and \textsf{stats} packages~\cite{R_pracma, RCoreTeam2020} were used. Signal processing steps are as follows:
\begin{enumerate}
	\item After the import of the data into \textsc{R}, the specially prepared \textsf{moving average} function is applied for the initial data filtering. The function detects possible NA's when the signal has a discontinued character and prevents filtering stops. Details can be found in the program~\cite{Kucharski2022_Rcode_neckle}. The moving average window with $80$ pixels was chosen for the presented result.
	\item The spline function \textsf{smooth.spline()} smoothed the filtered signal with the desired equivalent number of degrees of freedom (trace of the smoother matrix) $27$~\cite{DeBoor1972, Perperoglou2019, Quak_2016}. In this way, the signal is prepared for an extrema detection step.
	\item The \textsf{predict} function of the package \textsf{car} \textsc{R}~\cite{R_car} is used for the first derivative calculation of the smoothed signal.
	\item Using the \textsf{peaks} function of the \textsf{pracma} \textsc{R} package, the extrema of the first derivative are detected. The function returns the intensities as~$y$s.
	\item To find the extrema positions in pixels ($x$'s) the \textsf{which} \textsc{R} function is used for the signal edge detection (a shadow edge detection position).
	\item The estimated absolute difference between the positions of two points on the edge of the shadow defined as the width ($P$ $[$Pixels$]$ parameter) corresponds to the size of the measured object.  
\end{enumerate}
For preparing the manuscript graphs, the package \textsf{tikzDevice} was used~\cite{R_tikzDevice}.

\section{More result examples}\label{sec:app_more_data}
\begin{figure}[htbp!]
\centering\begin{subfigure}{.5\linewidth}
\centering\includegraphics[scale=0.66]{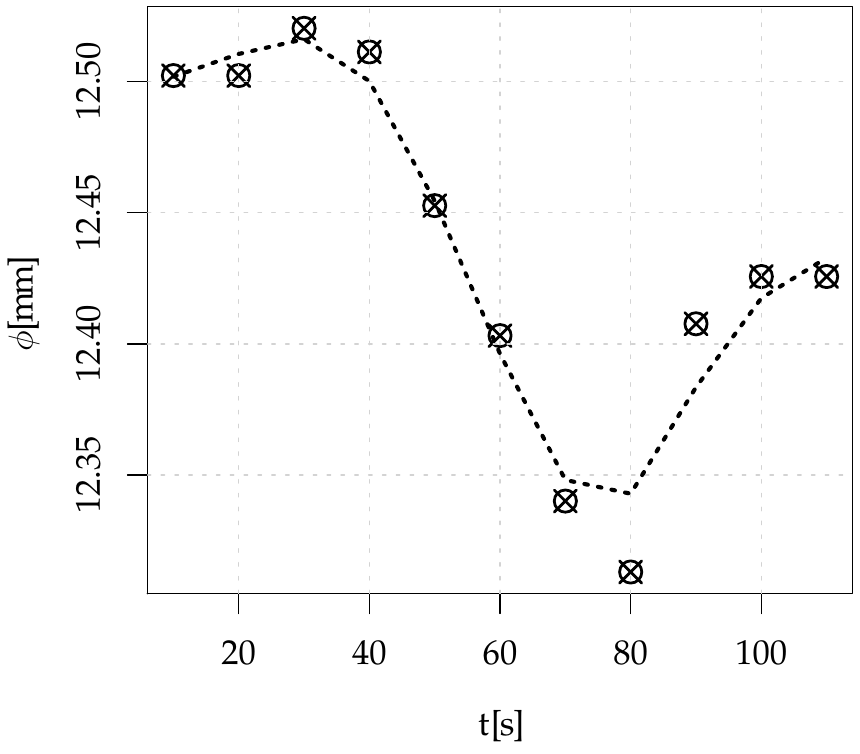}
                  \caption[]%
           {{\small}}
           \label{fig:pet_app_graph_1ch}
\end{subfigure}%
\hfill
\centering\begin{subfigure}{.5\linewidth}
		\centering\includegraphics[scale=0.66]{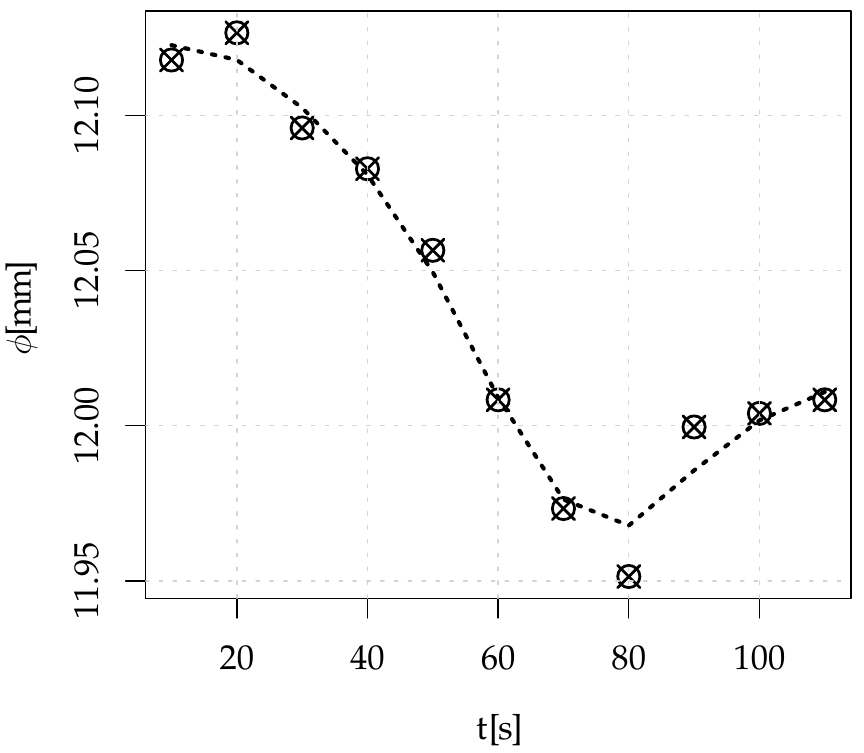}
           \caption[]%
           {{\small}}
           \label{fig:pet_app_graph_2ch}
\end{subfigure}
\vskip\baselineskip
\centering\begin{subfigure}{\linewidth}
\centering
			\includegraphics[scale=0.8]{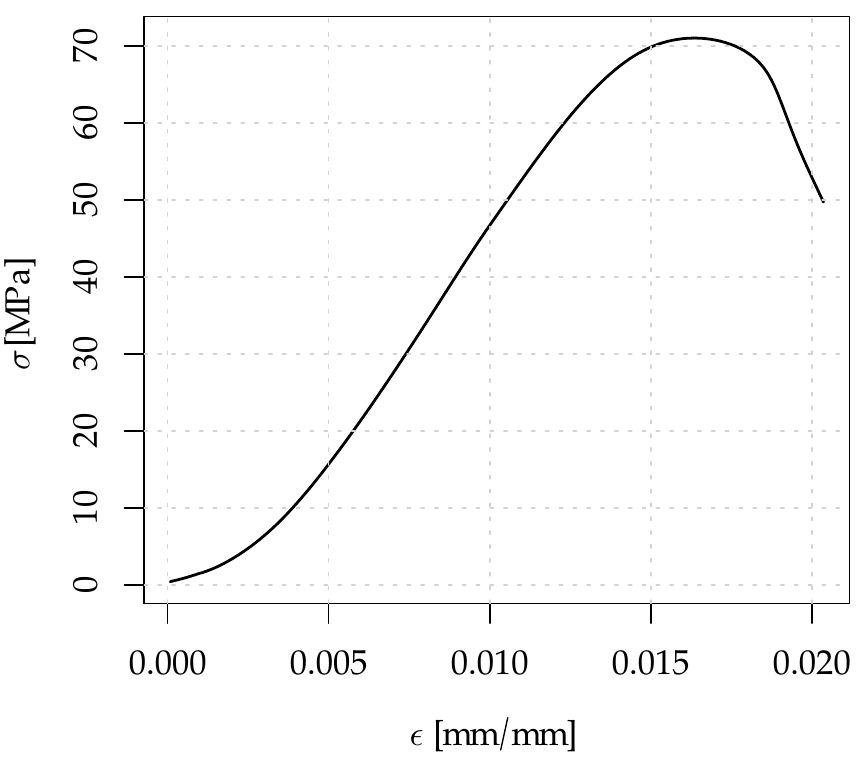}
           \caption[]%
           {{\small}}
           \label{fig:pet_app_stress}
\end{subfigure}
\caption[PET_app_results]
      {\small Example results for a solid PET rod with reduced detection points: (a) diameter change registered by channel $\#1$, (b) - diameter change registered by channel $\#2$, (c) - stress-strain curve obtained by the tensile machine. The points correspond to registered data; the dashed line corresponds to the b-spline fitting (degrees of freedom is $5$)}
          \label{fig:pet_app}
\end{figure}

\begin{figure}[htbp!]
\centering\begin{subfigure}{.5\linewidth}
           \centering
			\includegraphics[scale=0.66]{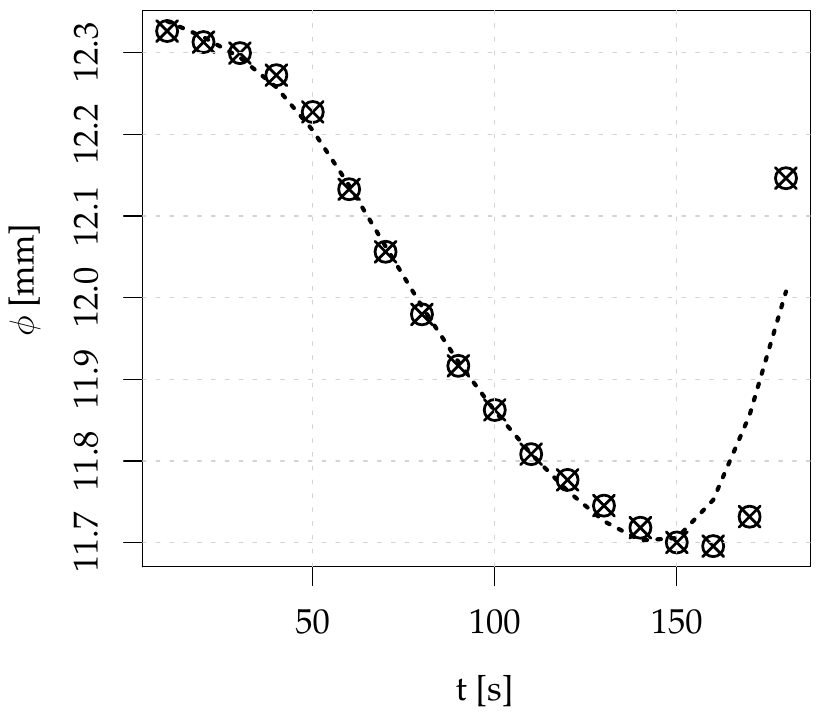}            
           \caption[]%
           {{\small}}
           \label{fig:pvdf3_app_graph_1ch}
\end{subfigure}%
\centering\begin{subfigure}{.5\linewidth}
      \centering
			\includegraphics[scale=0.66]{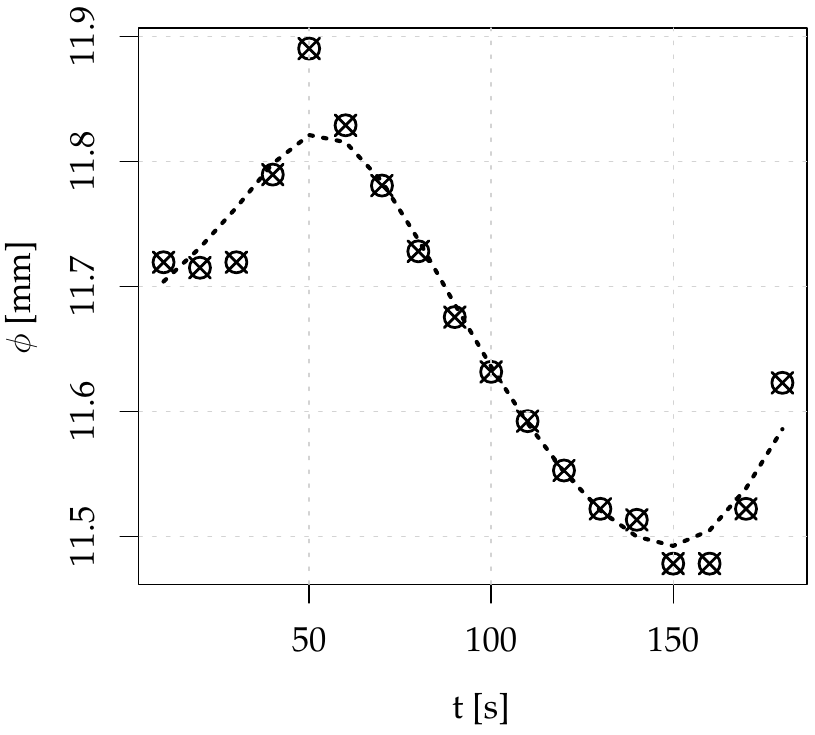}       
           \caption[]%
           {{\small}}
           \label{fig:pvdf3_app_graph_2ch}
\end{subfigure}
\vskip\baselineskip
\centering\begin{subfigure}{\linewidth}
\centering
			\includegraphics[scale=0.8]{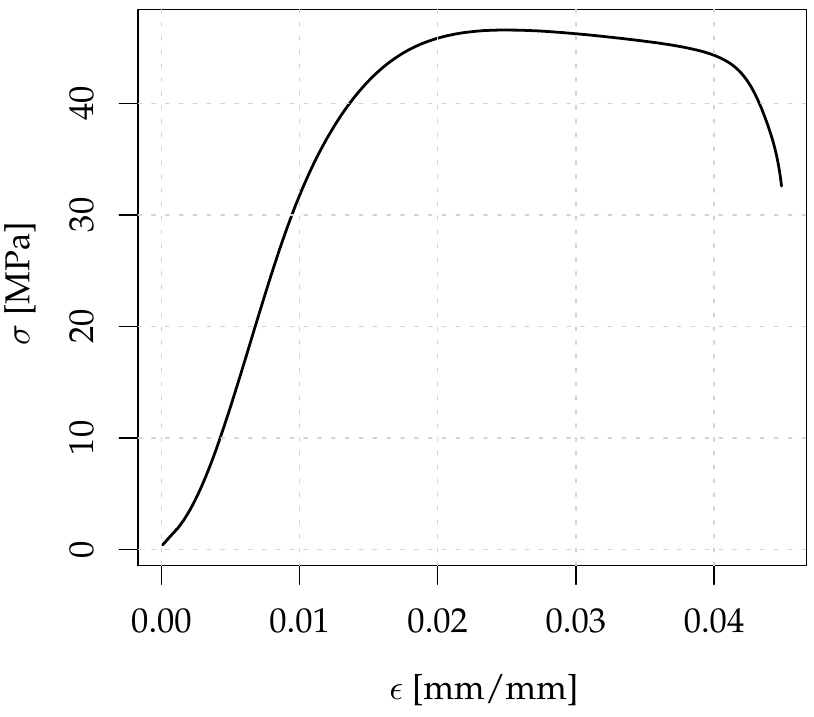}

           \caption[]%
           {{\small}}
           \label{fig:pvdf3_app_stress}
\end{subfigure}
\caption[PVDF_app_results]
      {\small Example results for a solid PVDF rod with reduced points: (a) - diameter change registered by channel $\#1$, (b) - diameter change registered by channel $\#2$, (c) - stress-strain curve obtained by the tensile machine. The points correspond to registered data; the dashed line corresponds to the b-spline fitting (degrees of freedom is $5$)}
      \label{fig:pvdf_app}           
\end{figure}

The measurement results with a reduced number of captured points are presented in figs.~\ref{fig:pet_app} and~\ref{fig:pvdf_app}. The growths of the detected diameters at the end of the tests correspond to the broken sample, which jumps out of the optical axis. In this way, clear necking is observed before a break after $80$ seconds of elongation for PET and after $150$ seconds for PVDF.

\section{Linear fitting of the calibration graphs}\label{sec:linear_fitting}
\begin{figure}[htbp!]
      \centering
      \begin{subfigure}[b]{0.485\textwidth}
          \centering
           \includegraphics[width=\textwidth]{./figs/tikz/7.pdf}
           \caption[]%
           {{\small}}
           
      \end{subfigure}
      \hfill
      \begin{subfigure}[b]{0.485\textwidth}
          \centering
          \includegraphics[width=\textwidth]{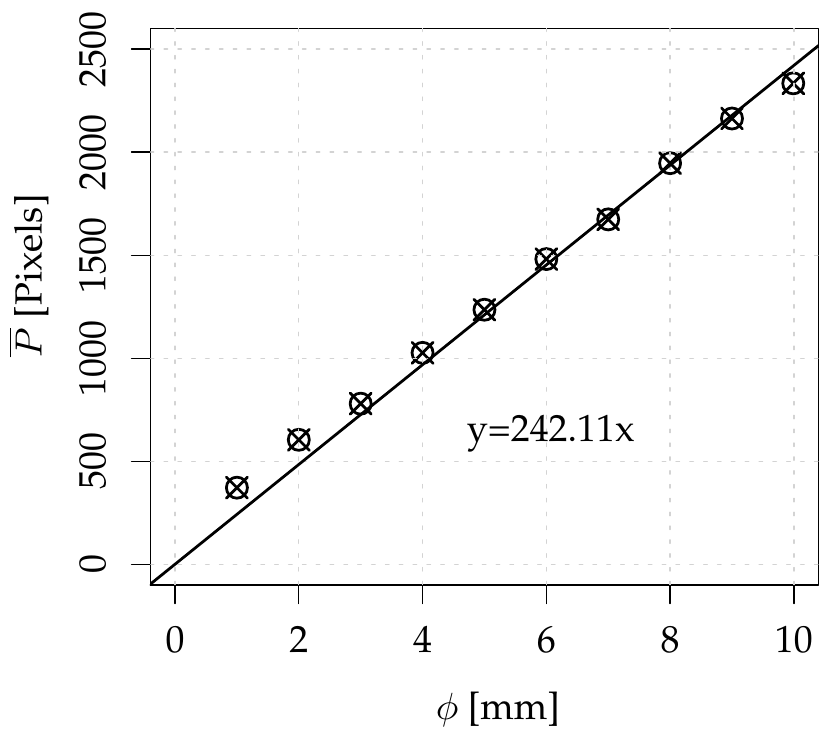}
         \caption[]%
          {{\small}}
          
      \end{subfigure}
      
      \begin{subfigure}[b]{0.485\textwidth}
          \centering
          \vskip\baselineskip
          \includegraphics[width=\textwidth]{./figs/tikz/8.pdf}
          \caption[]%
          {{\small}}
          
      \end{subfigure}
      \hfill
      \begin{subfigure}[b]{0.485\textwidth}
          \centering
          \includegraphics[width=\textwidth]{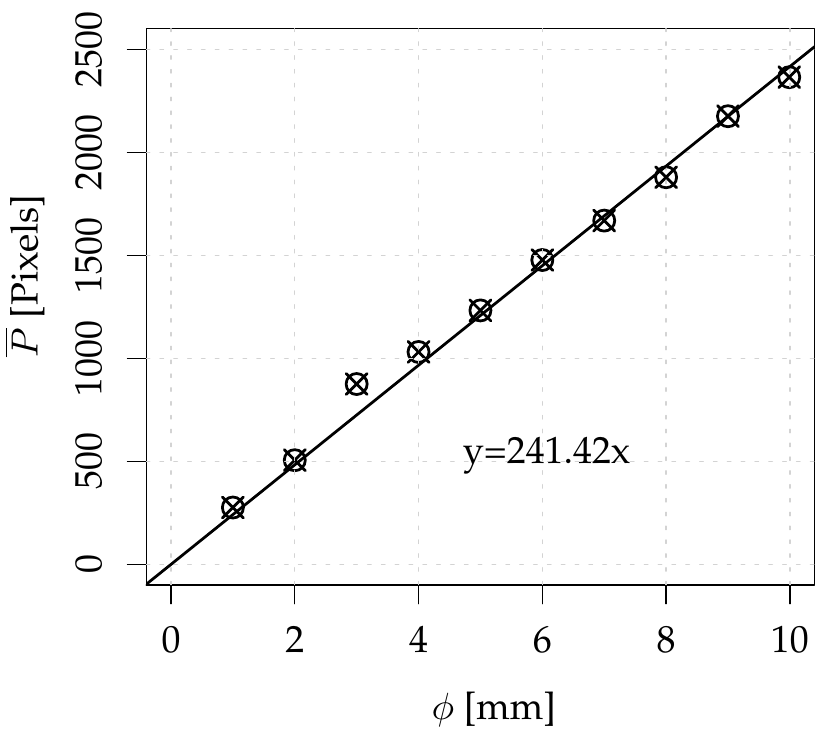}
          \caption[]%
          {{\small}}
    
      \end{subfigure}
      \caption[Calibration graphs]
      {\small 
      Comparison of calibration curves (linear regression $y=ax+b=\overline{P}=a\phi+b$) for two types of non-zero and zero intercepts fitting conditions for channel $\#1$: (a) -- fitted parameters: $a=221.96$, $b=141$, std. errors $\Delta a=2.47$, $\Delta b=15.36$, $p<0.001$; (b) -- fitted parameters: $a=241.11$, std. error $\Delta a=3.67$, $p<0.001$;  and for channel $\# 2$: (c) -- fitted parameters: $a=228.36$, $b=92.77$, std. errors $\Delta a=5.12$, $\Delta b=31.77$, $p<0.001$ and (d) -- fitted parameters: $a=241.42$, std. error $\Delta a=3.22$, $p<0.001$, respectively}
     \label{fig:calibration_graphs_supp}
  \end{figure}

For linear fitting of the models, the function $lm(formula, data)$ of \textsc{R} was used. An object of class $formula$ is a symbolic description of the model to be fitted. A $formula$ has an implicit intercept term of: $formula=y\sim x$. A fixed zero-intercept fitting $formula=y\sim 0+x$ can be applied to remove this. For the graph-based calibration method, the non-zero free intercept was applied to give a better approximation to the linear function later used to determine the diameter of samples in the $\phi=7-12$ mm range. The significant error of approximation is clearly to be seen, especially for $\phi=10$ mm of the zero intercept type fitting for both channels of the setup (see fig.~\ref{fig:calibration_graphs_supp})

\newpage
\section{A note on b-spline fitting}\label{sec:b-spline_supp}
The data's global nonlinear fitting has several issues, the essential being non-locality (proper function at a given value depends on data values far from that point). In spline fitting, instead of fitting a global polynomial, the range of $x$ is partitioned into smaller intervals, using an arbitrary number and position of points, $K$, also called knots. Splines are created using a polynomial with an interval between two successive knots. The spline has four parameters in each $K + 1$ region minus three constraints for each knot, giving $K+4$ degrees of freedom. For example, a spline with three knots ($K=3$) has seven degrees of freedom.

The \textsc{R} programming environment offers several packages that support b-spline fitting. For the data processing presented in the paper, the package \textsf{smooth.spline()} has been applied with specified degrees of freedom ($df$). The lower value of the degrees of freedom meant stronger data smoothing. The smoothing level is more dependent on the operator's choice based on the number of registered points rather than the error estimation procedure. For CCD signals, it was optimised as $df=27$, influenced the estimated precision of the system calibration $2\times \sigma < 5$ pixels.


\newpage
\bibliography{bib.bib}
\bibliographystyle{elsarticle-num}

\end{document}